\documentclass[aps,11pt,letterpaper]{revtex4}
\usepackage{amsmath,amssymb,mathptmx,slashed,bm}
\usepackage[pdftex]{graphicx}
\newcommand{\Lx}{\left(}
\newcommand{\Rx}{\right)}
\newcommand{\LB}{\left[}
\newcommand{\RB}{\right]}

\newcommand{\ep}{{\varepsilon}}
\newcommand{\eqn}[1]{Eq.\,(\ref{#1})}

\newcommand{\Sec}[1]{Section~\ref{#1}}
\newcommand{\app}[1]{Appendix~\ref{#1}}

\newcommand{\nlo}{{NLO}}
\newcommand{\nnlo}{{NNLO}}
\newcommand{\nnnlo}{{N${}^3$LO}}
\newcommand{\qcd} {{QCD}}

\newcommand{\cdr} {{\rm CDR}}
\newcommand{\hv} {{\rm HV}}
\newcommand{\fdh} {{\rm FDH}}
\newcommand{\dred} {{\rm DRED}}
\newcommand{\DR} {{\rm DR}}

\newcommand{\ms} {{\rm MS}}
\newcommand{\dr} {{\rm DR}}
\newcommand{\msbar} {{\overline{\ms}}}
\newcommand{\drbar} {{\overline{\dr}}}
\newcommand{\drhat} {{\widehat{\dr}}}

\newcommand{\amsbar} {{\alpha_{s}^{\msbar}}}
\newcommand{\amsbarpi} {{\Lx\frac{\alpha_{s}^{\msbar}}{\pi}\Rx}}

\newcommand{\betabar}[1] {{\beta_{#1}^{\msbar}}}

\newcommand{\betadrhat}[1] {{{\beta}_{#1}^{\drhat}}}
\newcommand{\betaedrhat}[2] {{{\beta}_{e,\,#1\,#2}^{\drhat}}}

\newcommand{\adrbar} {{\alpha_{s}^{\drbar}}}
\newcommand{\adrhat} {{\alpha_{s}^{\drhat}}}

\newcommand{\adrhatpi} {{\Lx\frac{\adrhat}{\pi}\Rx}}

\newcommand{\aedrhat} {{\alpha_{e}^{\drhat}}}

\newcommand{\etadrhat}[1] {{\eta_{#1}^{\drhat}}}

\newcommand{\etadrhatpi}[1] {{\Lx\frac{\etadrhat{{#1}}}{\pi}\Rx}}

\newcommand{\aedrhatpi} {{\Lx\frac{\aedrhat}{\pi}\Rx}}

\newcommand{\ket}[1]{\left| #1\right\rangle}

\newcommand{\CF}{C_F}
\newcommand{\CA}{C_A}
\newcommand{\nc}{N_c}

\begin{document}

\today

\bibliographystyle{apsrev}

\title{The Four Dimensional Helicity Scheme Beyond One Loop}

\author{William~B.~Kilgore}
\affiliation{Physics Department, Brookhaven National Laboratory,
  Upton, New York
  11973, USA.\\
  {\tt [kilgore@bnl.gov]} }

\begin{abstract}
I describe a procedure by which one can transform scattering
amplitudes computed in the four dimensional helicity scheme into
properly renormalized amplitudes in the 't~Hooft-Veltman scheme.  I
describe a new renormalization program, based upon that of the
dimensional reduction scheme and explain how to remove both finite and
infrared-singular contributions of the evanescent degrees of freedom
to the scattering amplitude.
\end{abstract}

\maketitle

\section{Introduction}
The Four Dimensional Helicity (\fdh)
scheme~\cite{Bern:1992aq,Bern:2002zk} is widely used for computing
\qcd\ corrections at next-to-leading order in perturbation theory.  It
is particularly convenient for use with the helicity method and the
techniques of generalized unitarity.  Unfortunately, as I have
recently shown~\cite{Kilgore:2011ta}, the \fdh\ is not a unitary
regularization scheme.  The standard renormalization
prescription~\cite{Bern:2002zk} fails to remove all of the ultraviolet
poles, leading to incorrect results at two loops and beyond.  Thus the
\fdh\ cannot be viewed as a regularization scheme in which one
can compute scattering amplitudes.  Instead, it should be looked upon
as a shortcut for obtaining scattering amplitudes in a unitary
regularization scheme.  Indeed, this is how the \fdh\ has always been
used at one-loop; final results have always been presented in the
't~Hooft-Veltman (\hv) scheme~\cite{'tHooft:1972fi} using the
prescription of Kunszt, et al.~\cite{Kunszt:1993sd} to transform the
\fdh\ scheme result, but it was not clear whether this conversion was
necessary or merely expedient, allowing one to match onto standard
definitions of the running coupling, etc.

It is now certain that one must convert the results of a calculation
in the \fdh\ scheme into results in a properly defined scheme.  A
first step in this direction was taken by Boughezal, et
al.~\cite{Boughezal:2011br}, who put forward a prescription for
constructing the correct counterterms for renormalization.  For
inclusive calculations, performed using the optical theorem, like
those considered in Refs.~\cite{Kilgore:2011ta,Boughezal:2011br}, such
a prescription is sufficient.  Experiments, however, measure
differential cross sections, and the power of the \fdh\ scheme
is that it facilitates the calculation of loop-level amplitudes,
giving access to the differential information they contain.  To make
use of the full amplitude, one must control of both the infrared and
ultraviolet structure.

In this paper, I will exploit the close relationship between the \fdh\
and the dimensional reduction (\dred)~\cite{Siegel:1979wq} schemes to
develop a prescription for transforming \fdh\ scheme amplitudes, which
may be easier to compute using unitarity methods, into \hv\ scheme
amplitudes that can actually be used in calculations.  The plan of the
paper is: In \Sec{section:regschemes} I will review the
regularization schemes that will be used; in \Sec{sec:irstructure} I
will review the infrared structure of \qcd\ amplitudes; in
\Sec{sec:fdhdef} I will define the \fdh\ scheme in terms of the \DR\
scheme, compute the anomalous dimensions that control the ultraviolet
and infrared structure of \DR\ scheme amplitudes through two loops and
specify the procedure for transforming \fdh\ scheme results into
\hv\ scheme amplitudes.

\section{Regularization schemes}
\label{section:regschemes}
All of the schemes that I will be working with are variations on
dimensional regularization~\cite{'tHooft:1972fi}, which specifies that
loop-momenta are to treated as $D_m = 4 - 2\,\ep$ dimensional.  In
dimensional regularization, the singularities (both ultraviolet and
infrared) that appear in four-dimensional calculations are transformed
into poles in the parameter $\ep$.  The ultraviolet poles are removed
through renormalization, while the infrared poles cancel when one
performs ``sufficiently inclusive'' calculations.

\subsection{The 't~Hooft-Veltman and conventional dimensional
  regularization schemes}
In the original dimensional regularization
scheme~\cite{'tHooft:1972fi}, the \hv\ scheme, observed states are
treated as four-dimensional, while internal states (both their momenta
and their spin degrees of freedom) are treated as $D_m$ dimensional.
Internal states include states that circulate inside of loop diagrams
as well as nominally external states that have infrared overlaps with
other nominally external states.  It turns out that one can treat
internal fermions as having exactly two degrees freedom, just as they
have in four dimensions, even though their momenta are $D_m$
dimensional, but massless internal gauge bosons must have $(D_m-2)$
spin degrees of freedom, while massive internal gauge bosons have
$(D_m-1)$.

The conventional dimensional regularization (\cdr)
scheme~\cite{Collins:Renorm} is closely related to the \hv\ scheme.
In the \cdr\ scheme, all states and momenta, both internal and
observed, are taken to be $D_m$ dimensional.  This often turns out to
be computationally more convenient, especially in infrared sensitive
theories like \qcd, since one set of rules governs all interactions.
Because the \hv\ and \cdr\ schemes handle ultraviolet singularities in
the same manner, their behavior under the renormalization group,
anomalous dimensions, running coupling, etc., are identical.

In the \hv\ and \cdr\ schemes, internal momenta are taken to be $D_m =
4 - 2\,\ep$ dimensional.  In general, $\ep$ is a complex number and
it's exact value is unimportant, but taking $\ep$ to be real and
positive (negative) is preferred by ultraviolet (infrared)
power-counting arguments.  It is important, however, that the
$D_m$-dimensional vector space in which momenta take values is {\it
larger} than the standard four-dimensional space-time.  This means
that the standard four-dimensional metric tensor $\eta^{\mu\nu}$ spans
a smaller space than the $D_m$ dimensional metric tensor, and the
four-dimensional Dirac matrices $\gamma^{0,1,2,3}$ form a subset of
the full $\gamma^\mu$.  These considerations are of particular
importance when considering chiral objects involving $\gamma_5$ and
the Levi-Civita tensor, but cannot be neglected when, as in the \hv\
scheme, one restricts observed states to be strictly four-dimensional.

\subsection{The dimensional reduction Scheme}
The \dred\ scheme was devised for application to supersymmetric
theories.  In supersymmetry, it is essential that the number of
bosonic degrees of freedom is exactly equal to the number of fermionic
degrees of freedom.  In the \dred\ scheme, the continuation to $D_m$
dimensions is taken as a {\it compactification} from four dimensions.
Thus, while space-time is taken to be four-dimensional and particles
have the standard number of degrees of freedom, momenta are
regularized dimensionally and span a $D_m$ dimensional vector space
which is {\it smaller} than four-dimensional space-time.

Because the Ward Identity only applies in the $D_m$ dimensional vector
space in which momenta are defined, the extra $2\,\ep$ spin degrees of
freedom of gauge bosons are not protected by the Ward Identity and
must renormalize differently than the $2-2\,\ep$ degrees of freedom
that are protected.  In supersymmetric theories, the supersymmetry
provides the missing part of the Ward Identity which demands that the
$2\,\ep$ spin degrees of freedom be treated as gauge bosons.  In
non-supersymmetric theories, however, they must be considered to be
distinct particles, with distinct couplings and renormalization
properties.  These extra degrees of freedom are referred to as
``$\ep$-scalars'' or as ``evanescent'' degrees of freedom.

Since the evanescent degrees of freedom are independent of the gauge
bosons, their self-couplings and their coupling to fermions are
independent of the gauge coupling and of one another.  The quartic
self-coupling splits into multiple independent terms; if the gauge
theory is $SU(2)$, there are two independent quartic self-couplings,
in $SU(3)$, there are three independent quartic self-couplings, and if
the gauge theory is $SU(N); N\ge4$, there are four independent quartic
self-couplings~\cite{Jack:1993ws}.  These new couplings run
differently from the gauge coupling under the renormalization group
and cannot consistently be identified with it.

Notwithstanding its semantic appeal, the insistence on a proper
compactification, so that $D_m\subset4$ in the \dred\ scheme, is
problematic when dealing with chiral theories~\cite{Siegel:1980qs}.
Chirality is a four-dimensional concept and one cannot consistently
define chiral operators in a vector space with fewer than four
dimensions.  One way around this is to adopt a hierarchy of vector
spaces $D_s\supset D_m\supset 4$ (where $D_m=4-2\,\ep$ and $D_s$ is
assigned the value $D_s=4$), as in the \fdh\ scheme (described below).
In such a scheme, chiral operators can be defined in the
four-dimensional subspace of $D_m$, just as they are in the \hv/\cdr\
schemes.  St\"ockinger and
Signer~\cite{Stockinger:2005gx,Signer:2008va} have long advocated that
this is the proper definition of the \dred\ scheme.  Aside from the
treatment of chiral operators, there are no important computational
distinctions between $D_m\supset4$ and $D_m\subset4$.  In this paper,
I will adopt the $D_m\supset4$ convention and refer to this variation
of dimensional reduction as the \DR\ scheme.

\subsection{The four dimensional helicity Scheme}
In the four-dimensional helicity scheme, one again defines a vector
space of dimensionality $D_m\supset4$ (again $D_m = 4-2\,\ep$), in
which loop momenta take values, and a still larger vector space
$D_s\supset D_m$, ($D_s=4$), in which internal spin degrees of freedom take
values.  Note that the relative numerical values of $D_s$, $D_m$ and
$4$ are not important.  What is important is that as vector spaces,
$D_s\supset D_m\supset 4$.

The \fdh\ scheme, like the \hv\ scheme, treats observed states as
four-dimensional, except, as in inclusive calculations, where there
are infrared overlaps among external states.  When infrared overlaps
occur, external states are taken to be $D_s$ dimensional.

As in the \dred\ scheme, spin degrees of freedom take values in a
vector space that is larger than that in which momenta take values.
It would seem, therefore, that the same remarks regarding the Ward
Identity and the conclusion that the $D_x = D_s - D_m$ dimensional
components of the gauge fields and their couplings must be considered
as distinct from the $D_m$ dimensional gauge fields and couplings
would apply.

That is not, however, how the \fdh\ scheme has been used.  All field
components in the $D_s$ dimensional space are treated as gauge fields
and no distinction is made between the couplings.  The reason for
doing this is to facilitate the use of helicity amplitudes in
conjunction with unitarity methods, the idea being to ``sew together''
(four dimensional) tree-level helicity amplitudes into loop-level
amplitudes.  While helicity methods can be used in the \cdr\
scheme~\cite{Kosower:1990ax}, they are most transparently and
compactly represented using four-dimensional external states.  Thus,
the \fdh\ scheme demands that the gluons circulating through loop
amplitudes have the same number of spin degrees of freedom as the
external gluons of helicity amplitudes.

Unfortunately, this framework fails to subtract all of the ultraviolet
poles~\cite{Kilgore:2011ta} and generates incorrect results.  The
evanescent couplings and degrees of freedom need to be renormalized
separately from their gauge boson counterparts, but there is no
mechanism within the \fdh\ for doing so.  The errors, however, are
only of order ${\cal O}(\ep^1)$ in \nlo\ calculations (which is the
level at which the \fdh\ has been used in practical calculations to
date) and therefore do not adversely affect those results.  At \nnlo\
the errors would be of order ${\cal O}(\ep^0)$ and at \nnnlo\
and beyond the errors would be singular in $\ep$.

\section{The infrared structure of \qcd\ amplitudes}
\label{sec:irstructure}

The infrared structure of \qcd\ amplitudes is governed by a set of
anomalous dimensions which allow one to predict, for any amplitude,
the complete infrared structure~\cite{Catani:1998bh,Sterman:2002qn}.
These anomalous dimensions are known completely, in both the massless
and massive cases for one and two loop amplitudes, and their
properties beyond the two-loop level are being actively
studied~\cite{Aybat:2006wq,
Aybat:2006mz,Mitov:2009sv,Becher:2009cu,Gardi:2009qi,Becher:2009qa,
Becher:2009kw,Gardi:2009zv,Dixon:2009ur,Mitov:2010xw}.

For a general $n$ parton scattering process, the set of partons is
labeled by ${\bf f} = \{f_i\}_{i=1\dots n}$.  In the formulation of
Refs.~\cite{Sterman:2002qn,Aybat:2006wq,Aybat:2006mz}, a renormalized
amplitude may be factorized into three functions: the jet function
${\cal J}_{\bf f}$, which describes the collinear dynamics of the
external partons that participate in the collision; the soft function
${\bf S_f}$, which describes soft exchanges between the external
partons; and the hard-scattering function $\ket{H_{\bf f}}$, which
describes the short-distance scattering process,
\begin{equation}
\ket{{\cal M}_{\bf f}\Lx p_i,\tfrac{Q^2}{\mu^2},\alpha_s(\mu^2),\ep\Rx} =
   {\cal J_{\bf f}}\Lx\alpha_s(\mu^2),\ep\Rx\
    {\bf S_{f}}\Lx p_i,\tfrac{Q^2}{\mu^2},\alpha_s(\mu^2),\ep\Rx\
    \ket{H_{\bf f}\Lx p_i,\tfrac{Q^2}{\mu^2},\alpha_s(\mu^2)\Rx}\,.
\label{eqn:qcdfact}
\end{equation}
The notation indicates that $\ket{H_{\bf f}}$ is a vector and 
${\bf S_f}$ is a matrix in color space~\cite{Catani:1996jh,
Catani:1997vz,Catani:1998bh}.  As with any factorization, there is
considerable freedom to move terms about from one function to the
others.  It is convenient~\cite{Aybat:2006wq,Aybat:2006mz} to define
the jet and soft functions, ${\cal J}_{\bf f}$ and ${\bf S_f}$, so
that they contain all of the infrared poles but only contain infrared
poles, while all infrared finite terms, including those at
higher-order in $\ep$, are absorbed into $\ket{H_{\bf f}}$.

\subsection{The jet function in the \hv/\cdr\ schemes}
The jet function ${\cal J}_{\bf f}$ is found to be the product of
individual jet functions ${\cal J}_{f_i}$ for each of the external
partons,
\begin{equation}
{\cal J}_{\bf f}\Lx\alpha_s(\mu^2),\ep\Rx = \prod_{i\in{\bf{f}}}\
   {\cal J}_{i}\Lx\alpha_s(\mu^2),\ep\Rx\,.
\end{equation}
Each individual jet function is naturally defined in terms of the
anomalous dimensions of the Sudakov form factor~\cite{Sterman:2002qn},
\begin{equation}
\begin{split}
\ln{\cal J}_i^\cdr\Lx\alpha_s(\mu^2),\ep\Rx &=
  -\amsbarpi\LB\frac{1}{8\,\ep^2}\gamma_{K\,i}^{(1)} +
  \frac{1}{4\,\ep}{\cal G}_i^{(1)}(\ep)\RB\\
  &\quad + \amsbarpi^2\left\{
   \frac{\betabar{0}}{8}\frac{1}{\ep^2}\LB\frac{3}{4\,\ep}\gamma_{K\,i}^{(1)}
    + {\cal G}_i^{(1)}(\ep)\RB -
    \frac{1}{8}\LB\frac{\gamma_{K\,i}^{(2)}}{4\,\ep^2} + \frac{{\cal
	G}_i^{(2)}(\ep)}{\ep}\RB  \right\} + \dots\,,
\label{eqn:logjetcdr}
\end{split}
\end{equation}
where
\begin{equation}
\begin{split}
    \gamma_{K\,i}^{(1)} &= 2\,C_i,\quad
    \gamma_{K\,i}^{(2)} = C_i\,K = C_i\LB \CA\Lx\frac{67}{18} -
    \zeta_2\Rx - \frac{10}{9}T_f\,N_f\RB,\quad C_q \equiv \CF,\quad
    C_g \equiv \CA,\\
  {\cal G}_{q}^{(1)} &= \frac{3}{2}\CF + \frac{\ep}{2}\CF\Lx8-\zeta_2\Rx, 
  \qquad {\cal G}_{g}^{(1)} = 2\,\betabar{0} - \frac{\ep}{2}\CA\,\zeta_2,\\
  {\cal G}_{q}^{(2)} &= \CF^2\Lx\frac{3}{16} - \frac{3}{2}\zeta_2 +
    3\,\zeta_3\Rx + \CF\,\CA\Lx\frac{2545}{432} + \frac{11}{12}\zeta_2
    - \frac{13}{4}\zeta_3\Rx - \CF\,T_f\,N_f\Lx\frac{209}{108} +
    \frac{1}{3}\zeta_2\Rx,\\
  {\cal G}_{g}^{(2)} &= 4\,\betabar{1} + \CA^2\Lx\frac{10}{27} -
    \frac{11}{12}\zeta_2 - \frac{1}{4}\zeta_3\Rx +
    \CA\,T_f\,N_f\Lx\frac{13}{27} + \frac{1}{3}\zeta_2\Rx +
    \frac{1}{2}\CF\,T_f\,N_f\,.
\label{eqn:cdranomdims}
\end{split}
\end{equation}

Although ${\cal G}_i$ and $\gamma_{K\,i}$ are defined through the
Sudakov form factor, they can be extracted from fixed-order
calculations~\cite{Gonsalves:1983nq,Kramer:1986sg,Matsuura:1987wt,
Matsuura:1989sm,Harlander:2000mg,Moch:2005id,Moch:2005tm}.
$\gamma_{K\,i}$ is the cusp anomalous dimension and represents a pure
pole term. The ${\cal G}_i$ anomalous dimensions contain terms at
higher order in $\ep$, but I only keep terms in the expansion that
contribute poles to $\ln\Lx{\cal J}_i\Rx$.  $\CF = (\nc^2-1)/(2\,\nc)$
denotes the Casimir operator of the fundamental representation of
SU($\nc$), while $\CA=\nc$ denotes the Casimir of the adjoint
representation. $N_f$ is the number of quark flavors and $T_f = 1/2$
is the normalization of the \qcd\ charge of the fundamental
representation.  $\zeta_n=\sum_{k=1}^{\infty}1/k^n$ represents the
Riemann zeta-function of integer argument $n$.

\subsection{The soft function in the \hv/\cdr\ schemes}
The soft function is determined entirely by the soft anomalous
dimension matrix ${\bm\Gamma}_{S_f}$,
\begin{equation}
\begin{split}
{\bf S_f}^\cdr\Lx p_i,\tfrac{Q^2}{\mu^2},\alpha_s(\mu^2),\ep\Rx
  &= 1 + \frac{1}{2\,\ep}\amsbarpi{\bm\Gamma}_{S_f}^{(1)} +
   \frac{1}{8\,\ep^2}\amsbarpi^2{\bm\Gamma}_{S_f}^{(1)}\times{\bm\Gamma}_{S_f}^{(1)}\\
  &\qquad- \frac{\betabar{0}}{4\,\ep^2}\amsbarpi^2{\bm\Gamma}_{S_f}^{(1)}
   + \frac{1}{4\,\ep}\amsbarpi^2{\bm\Gamma}_{S_f}^{(2)} + \dots\,.
\label{eqn:cdrsoft}
\end{split}
\end{equation}
In the color-space notation of
Refs.~\cite{Catani:1996jh,Catani:1997vz,Catani:1998bh}, the soft
anomalous dimension is given by~\cite{Aybat:2006wq,Aybat:2006mz}
\begin{equation}
{\bm\Gamma}_{S_f}^{(1)} = \frac{1}{2}\,\sum_{i\in{\bf f}}\ \sum_{j\ne i}
   {\bf T}_i\cdot{\bf T}_j\,\ln\Lx\frac{\mu^2}{-s_{ij}}\Rx,\qquad
 {\bm\Gamma}_{S_f}^{(2)} = \frac{K}{2}{\bm\Gamma}_{S_f}^{(1)}\,,
\label{eqn:cdrsoftanomdim}
\end{equation}
where $K = \CA\Lx67/18-\zeta_2\Rx - 10\,T_f\,N_f/9$ is the same constant
that relates the one- and two-loop cusp anomalous dimensions. The 
${\bf T}_i$ are the color generators in the representation of parton $i$
(multiplied by $(-1)$ for incoming quarks and gluons and outgoing
anti-quarks).

\section{The \fdh\ scheme at two loops}
\label{sec:fdhdef}
The failure of the \fdh\ scheme as a unitary regularization scheme
does not mean that it is of no value in computing higher-order
corrections beyond the next-to-leading order.  Even at \nlo, the \fdh\
scheme has always been used as a means of obtaining scattering
amplitudes in the \hv\ scheme.  There is no reason for that to change
at two loops.  The only difference is that one must recognize that the
\fdh\ scheme result is not a physical scattering amplitude, but only
an intermediate step toward obtaining one.

In formulating a prescription for converting \fdh\ scheme amplitudes
into \hv\ scheme amplitudes, the first problem to address, of course,
is that of renormalization.  One solution to the renormalization
problem, dubbed ``dimensional reconstruction,'' has been proposed by
Boughezal, et al.~\cite{Boughezal:2011br}.  The idea behind
dimensional reconstruction is that if one knows the the one-loop
behavior of an amplitude with arbitrary (integer) numbers of extra
spin dimensions (momenta are always $D_m$ dimensional) then the
correct two-loop amplitude can be determined from the renormalization
constants at different integer spin dimensions.  Note that it is a
basic assumption of dimensional reconstruction that when one is
computing a two-loop amplitude, the tree-level and one-loop terms that
contribute via renormalization are essentially trivial, and that there
is no appreciable cost to performing extra one-loop calculations if
doing so saves effort on the two-loop piece.  The transformations that
I will develop will also subscribe to this viewpoint.

While dimensional reconstruction is a completely valid approach to the
renormalization problem of the \fdh\ scheme, it does have some
drawbacks.  One drawback is that it appears that one must determine
new renormalization constants for each process at each order of
perturbation theory.  This is quite different from working within a
renormalizable theory, where the renormalization constants can be
determined in advance through the study of corrections to 1PI
Green functions.  A more serious drawback is that dimensional
reconstruction does not address the infrared structure of amplitudes
computed in the \fdh\ scheme.

It is certain that the infrared structure of \fdh\ scheme amplitudes
is not equal to that of \hv\ scheme amplitudes.  It is also clear from
optical theorem calculations~\cite{Kilgore:2011ta,Boughezal:2011br}
that once the renormalization problem is fixed, one could proceed with
\fdh\ scheme calculations because the infrared overlaps will sort
themselves out.  For differential calculations, one needs to know the
soft and collinear factorization properties of \fdh\ scheme amplitudes
in order to implement a subtraction scheme, but this has already been
worked out~\cite{Bern:1998sc,Bern:1999ry,Kosower:1999rx}.  The problem
is that all of the \fdh\ scheme amplitudes, real and virtual, contain
errors, though the structure of the errors is such that, after
renormalization, they cancel in the inclusive sum.  Even if one were
willing to live with such circumstances, one would still want to match
onto standard definitions of the running coupling and would have to
face the fact that parton distribution functions are only available in
the \cdr\ scheme.  A far better choice is to transform the result to a
framework like the \hv\ scheme that is known to be unitary and correct
and which can be easily connected to the parton distribution functions.

\subsection{The connection between the \fdh\ and \DR\ schemes}
In order to develop a rigorous set of rules for transforming \fdh\
amplitudes, it is necessary to define the \fdh\ scheme in terms of a
renormalizable scheme.  One can do this by exploiting the close
connection between the \fdh\ and \DR\ schemes.  When formulating the
\qcd\ Lagrangians in these schemes, one starts with the standard
Yang-Mills Lagrangian and then extends the fields into
$D_s$-dimensions.  In the \fdh\ scheme, one proceeds directly to the
development of Feynman rules involving the $D_s$-dimensional metric
tensor and Dirac matrices~\cite{Bern:1992aq,Bern:2002zk}.  In the \DR\
scheme, however, one first splits the gluon field into two independent
components, the $D_m$-dimensional gauge field and the
$D_x$-dimensional evanescent field~\cite{Jack:1993ws,Jack:1994bn,
Harlander:2006rj}.  The metric tensor and Dirac matrices also
decompose into orthogonal components.  Those new terms in the
Lagrangian that do not involve gauge fields are assigned new,
independent couplings.  The evanescent-quark-antiquark coupling is
given the value $g_e$ ($g_e^2 = 4\,\pi\,\alpha_e$) and the quartic
evanescent boson couplings are given values $\eta_{i\,,i=1,2,3}$,
where $\eta_1$ represents the quartic interaction that has the same
color flow as the quartic gluon coupling, while $\eta_{2,3}$ represent
the non-\qcd-like interactions.

Thus, all of the \DR\ scheme interactions are contained in those of
the \fdh\ scheme, they are simply not labeled by independent couplings
and evanescent Lorentz structures.The only exception to this statement
concerns the quartic evanescent boson couplings.  Because the
evanescent bosons are not protected by gauge symmetry, new quartic
interactions, with new color-flows among the evanescent bosons, are
generated by higher-order corrections which must be renormalized
independently of the \qcd-like quartic coupling that appears in the
classical Lagrangian.  In recognition of the fact that such terms will
occur, they are usually assigned independent couplings and added to
the effective \DR\ Lagrangian.  The \fdh\ scheme doesn't have such
couplings, but this does not present a problem.  The extra quartic
terms introduced to the \DR\ Lagrangian clean up the renormalization
procedure, but there is no reason that the couplings assigned to these
terms could not be chosen such that they do not contribute to a \DR\
scattering process until radiative corrections to the \qcd-like
interactions demand that they appear.

\subsection{The connection between the \DR\ and \cdr\ schemes}
From the formulation of the Lagrangians, one can also draw a
connection between the structure of the amplitudes in the \DR\ and
\cdr\ schemes.  In particular, the \DR\ scheme Lagrangian contains all
of the interactions that the \cdr\ scheme Lagrangian does, plus a host
of interactions involving the evanescent bosons.  This means that the
amplitudes in the \DR\ scheme can be partitioned into a part that is
identical to the \cdr\ scheme amplitude and a part that involves the
exchange of one or more evanescent bosons.  One need not consider the
case of external evanescent bosons since the \DR\ scheme
renormalization program ensures that such terms contribute to the
S-matrix at order $\ep$~\cite{Capper:1980ns,Jack:1993ws}.  The \DR\
scheme sub-amplitude that involves evanescent exchanges will
necessarily include a spin-sum over the evanescent degrees of freedom,
with the result that this sub-amplitude will be weighted by a factor
of $D_x = 2\,\ep$.  The only way that a term from the evanescent
sub-amplitude can make a finite (or singular) contribution to the full
amplitude is if it is weighted by ultraviolet or infrared poles.
Thus, the full evanescent contribution to an amplitude up to order
$\ep^0$ is part of the universal (ultraviolet or infrared) structure
of the amplitude, and is controlled by anomalous dimensions.  This
means that the evanescent contribution to an $n$-loop amplitude (that
is the part that is different from the \cdr\ amplitude) can be
determined entirely in terms of ultraviolet counterterms, jet and soft
functions and lower-order ($0$ to $(n-1)$-loop) hard-scattering
functions.  Thus, with a proper rearrangement of terms (the $\drhat$
scheme defined below), at any order $n$ the hard-scattering functions
in the two schemes are related by
\begin{equation}
   \ket{H_{\bf f}^{(n)}}_{\drhat} = \ket{H_{\bf f}^{(n)}}_{\hv} +
   {\cal O}(\ep).
\label{eqn:hvdrhat}
\end{equation}

\subsection{A new definition of the \fdh\ scheme}
Clearly, if one can draw a close connection between the \fdh\ and \DR\
schemes, one should be able to develop a prescription for the direct
transformation of an amplitude computed in the \fdh\ scheme to one
that is computed in the \hv\ scheme.  From the above considerations,
it is quite simple to state the connection.

The four-dimensional helicity scheme {\it is} the \DR\ scheme with two
extra conditions:
\begin{enumerate}
  \item
  External states are taken to be four dimensional.
  \item
  The evanescent couplings ($\alpha_e$ and $\eta_1$) are identified
  with $\alpha_s$.
\end{enumerate}

The first condition asserts the same distinction between the \fdh\ and
\DR\ schemes as exists between the \hv\ and \cdr\ schemes.  The
restriction to four-dimensional external states does not affect the
anomalous dimensions of the theory.  The ultraviolet counterterms and
the jet and soft functions are unchanged.  The only changes are to the
exact form of the finite hard-scattering matrix elements.  The
four-dimensional condition also forbids the appearance of external
evanescent states.  As mentioned before, the renormalization program
of the \DR\ scheme ensures that evanescent external states can only
contribute to the S-matrix at order $\ep$ or higher, so this
restriction is of no consequence.

The second condition is the one that violates unitarity and renders
the \fdh\ non-renormalizable.  The evanescent couplings need to be
renormalized differently than the \qcd\ coupling, but there is no
means of doing so once the couplings have been identified.  Therefore,
the \fdh\ can only be used to compute bare (unrenormalized) loop
amplitudes.

In the \DR\ scheme, on the other hand, one can determine the correct
ultraviolet counterterms, and the infrared counterterms needed to
remove the evanescent contribution, leaving the \hv\ scheme result.
By computing these counterterms in the \DR\ scheme and then
identifying the couplings, one obtains the counterterms needed to
shift from the \fdh\ to the \hv\ scheme.

\subsection{Ultraviolet counterterms for the \fdh}
When working within massless \qcd, it is only necessary to renormalize
the couplings.  It is common in dimensional reduction to determine
ultraviolet counterterms using modified minimal subtraction (this is
known as the $\drbar$ scheme), dropping evanescent terms, even if they
contain ultraviolet poles, because the factor of $D_x$ renders them
finite.  This procedure means that the renormalized coupling in the
$\drbar$ scheme, $\adrbar$ differs from the standard coupling
$\amsbar$ that appears in \hv/\cdr\ calculations by a finite
renormalization.  This finite renormalization corresponds precisely to
the $D_x/\ep$ terms that were dropped from the $\beta$-function.  My
goal is to remove all evanescent contributions, so I will include
$(D_x/\ep)^n$ terms in my definitions of the $\beta$-functions and
anomalous dimensions.  To distinguish it from the $\drbar$ scheme, I
will call this the $\drhat$ scheme.

Because there are so many independent couplings in the \DR\ scheme,
and because they mix under renormalization, the simple
$\beta_{0,1,2,\ldots}$ labeling of the $\msbar$ scheme is
insufficient.  Instead, I write,
\begin{equation}
\begin{split}
\betadrhat{} = \mu^2\frac{d}{d\,\mu^2}\frac{\adrhat}{\pi}
   &= -\Lx\ep\frac{\adrhat}{\pi}
   +\frac{\adrhat}{Z_{\adrhat}}\frac{\partial
      Z_{\adrhat}}{\partial\aedrhat}\,\betadrhat{e}
   +\frac{\adrhat}{Z_{\adrhat}}\frac{\partial
      Z_{\adrhat}}{\partial\etadrhat{i}}\,\betadrhat{\eta_i}\Rx 
    \Lx1 + \frac{\adrhat}{Z_{\adrhat}}
    \frac{\partial Z_{\adrhat}}{\partial\adrhat}\Rx^{-1}\\
   &= -\ep\frac{\adrhat}{\pi} - \sum_{i,j,k,l,m}\,\betadrhat{ijklm}
    \,\adrhatpi^{i}\,\aedrhatpi^{j}\,\etadrhatpi{1}^{k}
    \,\etadrhatpi{2}^{l}\,\etadrhatpi{3}^{m}\,.
\label{eqn:drhatbetadef1}
\end{split}
\end{equation}
Similar equations yield
\begin{equation}
\begin{split}
\betadrhat{e} = \mu^2\frac{d}{d\,\mu^2}\frac{\aedrhat}{\pi}
   &= -\ep\frac{\aedrhat}{\pi} - \sum_{i,j,k,l,m}\,\betadrhat{e,\,ijklm}
    \,\adrhatpi^{i}\,\aedrhatpi^{j}\,\etadrhatpi{1}^{k}
    \,\etadrhatpi{2}^{l}\,\etadrhatpi{3}^{m}\,,\\
\betadrhat{\eta_s} = \mu^2\frac{d}{d\,\mu^2}\frac{\etadrhat{s}}{\pi}
   &= -\ep\,\frac{\etadrhat{s}}{\pi} - \sum_{i,j,k,l,m}\,\betadrhat{s,\,ijklm}
    \,\adrhatpi^{i}\,\aedrhatpi^{j}\,\etadrhatpi{1}^{k}
    \,\etadrhatpi{2}^{l}\,\etadrhatpi{3}^{m}\,.
\label{eqn:drhatbetadef2}
\end{split}
\end{equation}
The values of the coefficients through three loops (for $\betadrhat{}$
and $\betadrhat{e}$) are given in \app{sec:betacoeffs}.  Note that
with the rearrangement of the evanescent contributions, the terms in
$\betadrhat{}$ that are not proportional to $D_x$ are identical to the
coefficients of the $\beta$-function in the $\msbar$ scheme.  This
indicates that the renormalized coupling of the $\drhat$ scheme
coincides with that of the $\msbar$ scheme.

The ultraviolet counterterms for \fdh\ amplitudes are computed as
follows.  First, one computes the lower loop amplitudes in the \DR\
scheme and then expands the bare couplings in terms of the
renormalized couplings using the $\beta$-functions of the $\drhat$
scheme.  Finally, the evanescent couplings are identified with the
\qcd\ coupling and the factors of $D_x$ are evaluated ($D_x=2\,\ep$).

\begin{equation}
\ket{{\cal M}(\alpha_s)}^{\rm CT}_{\fdh} = 
  \left.\ket{{\cal M}(\alpha_s,\alpha_e,\eta_1)}^{\rm CT}_{\drhat}
  \right|_{\genfrac{}{}{0pt}{}
  {\alpha_e,\eta_1\to\alpha_s}{D_x\to2\,\ep\hfill}}
\label{eqn:fdhrenorm}
\end{equation}
This will remove all of the ultraviolet terms, including the
evanescent terms that appear to be finite because of the factor of
$D_x$.

\subsection{The infrared structure of the \DR\ scheme}
The next step is to remove the unwanted evanescent component of the
infrared structure of \fdh\ scheme amplitudes.  As with the
ultraviolet counterterms, the terms to be removed can be identified by
studying the structure of \DR\ scheme amplitudes.  The basic form of
the infrared structure in the \DR\ scheme is the same as in \hv/\cdr,
but the anomalous dimensions receive evanescent corrections.  In
addition, there are new ${\cal G}$ anomalous dimensions that depend on
the evanescent couplings.  Through two-loops, the corrections and new
anomalous dimensions depend only on the fermion-evanescent coupling,
not the quartic evanescent couplings.  Furthermore, because the
evanescent couplings are not gauge couplings, there are no new
counterparts to the cusp or soft anomalous dimensions, which are
associated with the exchange of gauge bosons.

I have determined the values of the infrared anomalous dimensions in
the \DR\ scheme by the direct calculation of two-loop amplitudes.  I
first determine the anomalous dimensions for external quarks from the
Drell-Yan amplitude.  I then obtain the anomalous dimensions for
external gluons from the $q\overline{q}\to\ g\gamma$
amplitude~\cite{Anastasiou:2001sv,Anastasiou:2002zn,Glover:2003cm}.
In principle, it would be easier to extract the gluon jet function by
calculating the amplitude for $g\,g \to\ H$, but the Higgs - gluon
coupling is governed by a set of effective operators generated by
integrating out the top quark.  This system, involving operator mixing
and higher-order corrections to the Wilson coefficients, has been
studied to high order in the \cdr\ scheme~\cite{Chetyrkin:1997iv,
Chetyrkin:1998un}, but not in the non-supersymmetric \DR\ scheme.

The calculations of the infrared anomalous dimensions as well as the
wave-function and vertex corrections used to extract the
$\beta$-functions were all calculated within the same framework.  The
Feynman diagrams were generated with QGRAF~\cite{Nogueira:1993ex} and
the symbolic algebra program FORM~\cite{Vermaseren:2000nd} was used to
implement the Feynman rules and perform algebraic manipulations to
reduce the result to a set of Feynman integrals and their
coefficients.  The method of Ref.~\cite{Davydychev:1997vh} was used to
reduce the calculation of the vertex corrections to propagator
integrals.  The full set of Feynman integrals was reduced to master
integrals using the program REDUZE-2~\cite{vonManteuffel:2012yz}.
REDUZE-2 offers significant improvements over the previous
version~\cite{Studerus:2009ye} and was particularly effective at
reducing the non-planar double-box integrals that contribute to the
$q\overline{q}\to\ g\gamma$ amplitude.  All of the master integrals
needed for these calculations are known in the literature~\cite{
Chetyrkin:1980pr,Kazakov:1983ns,Gehrmann:2005pd,Smirnov:1999gc,
Anastasiou:2000kp,Tausk:1999vh,Anastasiou:2000mf}.

The jet function in the \DR\ scheme takes the form,
\begin{equation}
\begin{split} 
\ln\widehat{\cal J}_{i}^\DR\Lx\alpha_s(\mu^2),\alpha_e(\mu^2),\ep\Rx =
 & -\amsbarpi\LB\frac{1}{8\,\ep^2}\hat{\gamma}_{K\,i}^{(1)} +
  \frac{1}{4\,\ep}\widehat{\cal G}_i^{(1)}(\ep)\RB
  -\aedrhatpi \frac{\widehat{\cal G}_{i,e}^{(0,1)}(\ep)}{4\,\ep}\\
  & + \amsbarpi^2 \LB\frac{\betadrhat{20}}{8}
   \frac{1}{\ep^2}\Lx\frac{3}{4\,\ep}\hat{\gamma}_{K\,i}^{(1)}
   + \widehat{\cal G}_i^{(1)}(\ep)\Rx
   - \frac{1}{8}\Lx\frac{\hat{\gamma}_{K\,i}^{(2)}}{4\,\ep^2}
   + \frac{\widehat{\cal G}_i^{(2)}(\ep)}{\ep}\Rx  \RB\\
  & + \amsbarpi\aedrhatpi\frac{1}{8}\LB
    \frac{\betaedrhat{1}{1}\,\widehat{\cal G}_{i,e}^{(0,1)}(\ep)}{\ep^2}
  - \frac{\widehat{\cal G}_{i,e}^{(1,1)}(\ep)}{\ep} \RB\\
  & + \aedrhatpi^2\frac{1}{8} \LB
   \frac{\betaedrhat{0}{2}\,\widehat{\cal G}_{i,e}^{(0,1)}(\ep)}{\ep^2}
   - \frac{\widehat{\cal G}_{i,e}^{(0,2)}(\ep)}{\ep}
   \RB + \dots\,,
\label{eqn:logjetDR}
\end{split}
\end{equation}
where the anomalous dimensions in the $\drhat$ scheme are
\begin{equation}
\begin{split}
    \hat\gamma_{K\,i}^{(1)} &= 2\,C_i,\quad
    \hat\gamma_{K\,i}^{(2)} = C_i\,\hat{K} = C_i\LB \CA\Lx\frac{67}{18} -
    \zeta_2\Rx - \frac{10}{9}T_f\,N_f - \frac{2}{9}D_x\,\CA\RB,\quad
    C_q \equiv \CF,\quad C_g \equiv \CA,\\
  \widehat{\cal G}_{q}^{(1)} &= \frac{3}{2}\CF + \frac{\ep}{2}\CF\Lx8-\zeta_2\Rx, 
  \hskip133pt \widehat{\cal G}_{g}^{(1)} = 2\,\betadrhat{20} - \frac{\ep}{2}\CA\,\zeta_2,\\
  \widehat{\cal G}_{q,e}^{(0,1)} &= - \frac{1}{4}D_x\,\CF\,,
    \hskip195pt \widehat{\cal G}_{g,e}^{(0,1)} = 0\,,\\
  \widehat{\cal G}_{q}^{(2)} &= \CF^2\Lx\frac{3}{16} - \frac{3}{2}\zeta_2 +
    3\,\zeta_3\Rx + \CA\,\CF\Lx\frac{2545}{432} + \frac{11}{12}\zeta_2
    - \frac{13}{4}\zeta_3\Rx - \CF\,T_f\,N_f\Lx\frac{209}{108} +
    \frac{1}{3}\zeta_2\Rx\\
    & - D_x\,\CA\,\CF\Lx\frac{311}{864} + \frac{1}{24}\zeta_2\Rx\,,\\
  \widehat{\cal G}_{g}^{(2)} &= 4\,\betadrhat{30} + \CA^2\Lx\frac{10}{27} -
    \frac{11}{12}\zeta_2 - \frac{1}{4}\zeta_3\Rx +
    \CA\,T_f\,N_f\Lx\frac{13}{27} + \frac{1}{3}\zeta_2\Rx +
    \frac{1}{2}\CF\,T_f\,N_f\\
    & + D_x\,\CA^2\Lx\frac{7}{54} + \frac{1}{24}\zeta_2\Rx\,,\\
  \widehat{\cal G}_{q,e}^{(1,1)} &= D_x\Lx-\frac{11}{16}\CA\,\CF
       + \frac{1}{4}\CF^2 + \frac{1}{4}\CF^2\,\zeta_2\Rx\,,\hskip95pt
  \widehat{\cal G}_{g,e}^{(1,1)} =2\,\betadrhat{21}\,,\\
  \widehat{\cal G}_{q,e}^{(0,2)} &= \frac{3}{16}D_x\,\CF\,T_f\,N_f\,,\hskip177pt
  \widehat{\cal G}_{g,e}^{(0,2)} =0\,,\\
    \betadrhat{20} &= \frac{11}{12}\CA - \frac{1}{6}N_f
      - \frac{1}{24}D_x\,\CA\,,\\
    \betadrhat{30} &= \frac{17}{24}\CA^2 - \frac{5}{24}\CA\,N_f
      - \frac{1}{8}\CF\,N_f - \frac{7}{48}D_x\,\CA^2\,,\hskip 60pt
    \betadrhat{21} = \frac{1}{16}D_x\,\CF\,N_f\,,\\
    \betaedrhat{0}{2} &= \frac{1}{2}\CA - \CF - \frac{1}{4}N_f
           - \frac{1}{4}D_x\,\Lx \CA-\CF\Rx\,,\hskip77pt
    \betaedrhat{1}{1} = \frac{3}{2}\CF\,.
\label{eqn:DRanomdims}
\end{split}
\end{equation}
Note that the \qcd\ coupling is $\amsbar$, the same coupling used in
\hv/\cdr\ calculations.  Since I extract the anomalous dimensions from
amplitude calculations, I cannot separate the order $\ep$ part of the
one-loop $\widehat{\cal G}$ anomalous dimensions, which contributes at
two-loops when multiplied by a $\beta$-function coefficient, from the
pure two-loop $\widehat{\cal G}$ anomalous dimensions.  This merely
constitutes a rearrangement of terms and does not affect the
prediction of the infrared structure.

The soft function changes very little in going to the \DR\ scheme.
This is because evanescent exchanges do not add new soft anomalous
dimensions, they only add corrections to the existing terms.

\begin{equation}
\begin{split}
\widehat{\bf S_f}^\DR\Lx p_i,\tfrac{Q^2}{\mu^2},\alpha_s(\mu^2),\ep\Rx &= 
   1 + \frac{1}{2\,\ep}\amsbarpi\widehat{\bm\Gamma}_{S_f}^{(1)} +
   \frac{1}{8\,\ep^2}\amsbarpi^2\widehat{\bm\Gamma}_{S_f}^{(1)}
   \times\widehat{\bm\Gamma}_{S_f}^{(1)}\\
  &\qquad- \frac{\betadrhat{20}}{4\,\ep^2}\amsbarpi^2\widehat{\bm\Gamma}_{S_f}^{(1)}
   + \frac{1}{4\,\ep}\amsbarpi^2\widehat{\bm\Gamma}_{S_f}^{(2)}\,,
\label{eqn:DRsoft}
\end{split}
\end{equation}
\begin{equation}
\widehat{\bm\Gamma}_{S_f}^{(1)} = \frac{1}{2}\,\sum_{i\in{\bf f}}\ \sum_{j\ne i}
   {\bf T}_i\cdot{\bf T}_j\,\ln\Lx\frac{\mu^2}{-s_{ij}}\Rx,\qquad
 \widehat{\bm\Gamma}_{S_f}^{(2)}
    = \frac{\hat{K}}{2}\widehat{\bm\Gamma}_{S_f}^{(1)}\,,
\label{eqn:DRsoftanomdim}
\end{equation}
where $\hat{K} = \CA\Lx67/18-\zeta_2\Rx - 10/9\,T_f\,N_f -
2/9\,D_x\,\CA$ is again the same constant that relates the one- and
two-loop cusp anomalous dimensions, this time in the $\drhat$ scheme.

\subsection{Transforming \fdh\ amplitudes into \hv\ amplitudes}
I have now assembled all of the pieces needed to convert bare
amplitudes computed in the \fdh\ scheme into renormalized amplitudes
in the \hv\ scheme. To obtain an $n$-loop amplitude in the \hv\
scheme, one needs
\begin{enumerate}
  \item
  The bare $n$-loop amplitude in the \fdh\ scheme.
  \item
  The renormalized $m$-loop amplitudes ($m\in\{0,\dots,n-1\}$) to order
$\ep^{2\,(n-m)}$ in the \hv\ scheme.
  \item
  The jet and soft functions to order $n$ in the \hv\ scheme.
  \item
  The renormalized $m$-loop amplitudes ($m\in\{0,\dots,n-1\}$) to order
$\ep^{2\,(n-m)}$ in the $\drhat$ scheme.
  \item
  The jet and soft functions to order $n$ in the $\drhat$ scheme.
\end{enumerate}
Note that computing the $n$-loop squared amplitude to order $\ep^0$
already required the higher-order in $\ep$ contributions to the
lower-loop amplitudes in the \hv\ scheme.  The conversion procedure
requires them in the $\drhat$ scheme as well.

The first step is to expand \eqn{eqn:qcdfact} by orders of $\alpha_s$,
\begin{equation}
\begin{split}
  \ket{{\cal M}^{(n)}}_\hv &= \sum_{i=0}^n\,
    \LB{\cal J}\otimes{\bf S}\RB^{(i)}\ket{{\cal H}^{(n-i)}}_\hv\\
  \ket{{\cal M}^{(n)}}_\drhat &= \sum_{i=0}^n\,
    \LB\widehat{\cal J}\otimes\widehat{\bf S}\RB^{(i)}\ket{{\cal H}^{(n-i)}}_\drhat\\
\end{split}
\end{equation}
I now define the ``renormalized'' \fdh\ scheme amplitude as
\begin{equation}
\ket{{\cal M}^{(n)}}_\fdh = \ket{{\cal M}^{(n)}}_\fdh^{\rm Bare} +
    \ket{{\cal M}^{(n)}}^{\rm CT}_{\fdh} =
    \left.\ket{{\cal M}^{(n)}}_\drhat\right|_{\genfrac{}{}{0pt}{}
  {\alpha_e,\eta_1\to\alpha_s}{D_x\to2\,\ep\hfill}}\,.
\end{equation}
From this I find that
\begin{equation}
\left.\ket{{\cal H}^{(n)}}_\drhat\right|_{\genfrac{}{}{0pt}{}
  {\alpha_e,\eta_1\to\alpha_s}{D_x\to2\,\ep\hfill}}
   = \ket{{\cal M}^{(n)}}_\fdh
   - \sum_{i=1}^n\,\LB\widehat{\cal J}\otimes\widehat{\bf S}\RB^{(i)}
      \left.\ket{{\cal H}^{(n-i)}}_\drhat\right|_{\genfrac{}{}{0pt}{}
  {\alpha_e,\eta_1\to\alpha_s}{D_x\to2\,\ep\hfill}}\,.
\end{equation}
Finally, using \eqn{eqn:hvdrhat}, I obtain
\begin{equation}
 \ket{{\cal H}^{(n)}}_\hv = \ket{{\cal M}^{(n)}}_\fdh^{\rm Bare} +
    \ket{{\cal M}^{(n)}}^{\rm CT}_{\fdh}
   - \sum_{i=1}^n\,\LB\widehat{\cal J}\otimes\widehat{\bf S}\RB^{(i)}
      \left.\ket{{\cal H}^{(n-i)}}_\drhat\right|_{\genfrac{}{}{0pt}{}
  {\alpha_e,\eta_1\to\alpha_s}{D_x\to2\,\ep\hfill}} + {\cal O}(\ep)\,.
\end{equation}
The infrared structure of the \hv\ scheme amplitude can be extracted
from $\ket{{\cal M}^{(n)}}_\fdh^{\rm Bare}$ in a similar way or
constructed directly in terms of the lower order hard scattering
matrix elements and the jet and soft functions.

Let me now write out explicitly the transformation of a one-loop bare
amplitude in the \fdh\ scheme, involving $n_q$ quarks and anti-quarks
and $n_g$ gluons, into a renormalized one-loop amplitude in the \hv\
scheme.  Starting with
\begin{equation}
 \ket{{\cal H}^{(1)}}_\hv = \ket{{\cal M}^{(1)}}_\fdh^{\rm Bare} +
    \ket{{\cal M}^{(1)}}^{\rm CT}_{\fdh}
   - \LB\widehat{\cal J}+\widehat{\bf S}\RB^{(1)}
      \left.\ket{{\cal H}^{(0)}}_\drhat\right|_{\genfrac{}{}{0pt}{}
  {\alpha_e,\eta_1\to\alpha_s}{D_x\to2\,\ep\hfill}} + {\cal O}(\ep)\,,
\end{equation}
I add in the infrared parts of the \hv\ amplitude (note that the
one-loop soft functions of the \hv\ and $\drhat$ scheme are identical)
to obtain
\begin{equation}
\begin{split}
 \ket{{\cal M}^{(1)}}_\hv &= \ket{{\cal M}^{(1)}}_\fdh^{\rm Bare} -
    \amsbarpi\Lx\frac{n_q+n_g-2}{2\,\ep}\betadrhat{20}\Rx
    \ket{{\cal H}^{(0)}}_{\hv}\\
  &\hskip90pt + \Lx{\cal J}^{(1)} - \widehat{\cal J}^{(1)}\Rx_{\genfrac{}{}{0pt}{}
     {\alpha_e,\eta_1\to\alpha_s}{D_x\to2\,\ep\hfill}}
      \ket{{\cal H}^{(0)}}_\hv + {\cal O}(\ep)\\
    &= \ket{{\cal M}^{(1)}}_\fdh^{\rm Bare} -
    \amsbarpi\frac{n_q+n_g-2}{2\,\ep}\betabar{0}\ket{{\cal H}^{(0)}}_{\hv}\\
    &+ \amsbarpi\Lx\frac{n_q+n_g-2}{24}\CA - \frac{n_q}{8}\CF
      - \frac{n_g}{24}\CA\Rx\ket{{\cal H}^{(0)}}_\hv
      + {\cal O}(\ep)
\end{split}
\end{equation}
The first line is just the bare one-loop amplitude with standard
$\msbar$ ultraviolet counterterm, while the second line is the finite
shift, broken into ultraviolet, infrared $n_q$ and infrared $n_g$
pieces, identified by Kunszt, et al.~\cite{Kunszt:1993sd}.  Beyond one
loop, the transformations are not so simple and involve the structure
of the amplitudes in addition to the identities of the external
states.

\section{Conclusion}
In this paper, I have described a procedure for transforming bare loop
amplitudes computed in the four dimensional helicity scheme into
renormalized amplitudes in the 't~Hooft-Veltman scheme.  One of the
simplifying features of the \fdh, the treatment of the evanescent
states as if they were gluons, renders the scheme non-renormalizable.
Nevertheless, the \fdh\ can be defined in terms of a renormalizable
scheme, a variant of the dimensional reduction scheme.  Through this
connection to the \DR\ scheme, I have shown that the differences
between amplitudes calculated in the \fdh\ scheme and the \hv\ scheme
(up to order $\ep^-$) are either ultraviolet or infrared in origin and
are therefore part of the universal structure of the amplitude which
is controlled by anomalous dimensions.  By computing these anomalous
dimensions in the $\drhat$ scheme, defined above, through two loops, I
provide concrete formul\ae\ for the transformation of the amplitudes.

The utility of such transformations lies in the close connection
between the \fdh\ scheme and the techniques of generalized unitarity
and the helicity method.  These techniques are a natural fit for the
\fdh\ scheme, but the results need to be transformed into a
renormalizable scheme so that they can be used in practical
calculations.  With the procedures described in this paper, such
transformations can be performed.

\vskip20pt

\paragraph*{Acknowledgments:}
This research was supported by the U.S.~Department of Energy under
Contract No.~DE-AC02-98CH10886.

\appendix
\section{$\drhat$ Scheme $\beta$-functions}
\label{sec:betacoeffs}
The non-vanishing coefficients for $\betadrhat{}$ through three loops
are:
\begin{equation}
\begin{split}
    \betadrhat{20} &= \frac{11}{12}\CA - \frac{1}{6}N_f
      - \frac{1}{24}D_x\,\CA\,,\\
    \betadrhat{30} &= \frac{17}{24}\CA^2 - \frac{5}{24}\CA\,N_f
      - \frac{1}{8}\CF\,N_f - \frac{7}{48}D_x\,\CA^2\,,\hskip 60pt
    \betadrhat{21} = \frac{1}{16}D_x\,\CF\,N_f\\
    \betadrhat{40} &= \frac{2857}{3456}\CA^3
      - \frac{1415}{3456}\CA^2\,N_f - \frac{205}{1152}\CA\,\CF\,N_f
      + \frac{1}{64}\CF^2\,N_f + \frac{79}{3456}\CA\,N_f^2
      + \frac{11}{576}\CF\,N_f^2\\
     & + D_x\Lx- \frac{2749}{6912}\CA^3 + \frac{13}{432}\CA^2\,N_f
       + \frac{23}{2304}\CA\,\CF\,N_f\Rx + \frac{145}{13824}D_x^2\,\CA^3\\
   \betadrhat{31} & = D_x\Lx\frac{5}{256}\CA^2\,N_f
       + \frac{7}{32}\CA\,\CF\,N_f + \frac{3}{128}\CF^2\,N_f\Rx\\
   \betadrhat{22} &= D_x\Lx- \frac{1}{64}\CA^2\,N_f
       + \frac{7}{128}\CA\,\CF\,N_f - \frac{3}{64}\,\CF^2\,N_f
       + \frac{1}{256}\CA\,N_f^2 - \frac{7}{256}\CF\,N_f^2\Rx\\
     & + D_x^2\Lx\frac{1}{256}\CA^2\,N_f
       - \frac{5}{256}\CA\,\CF\,N_f\Rx\\
   \betadrhat{30100} &= \frac{27}{512}D_x\Lx1 - D_x\Rx\,, \hskip 30pt
   \betadrhat{30010} = -\frac{45}{126}D_x\Lx2 + D_x\Rx\,,
   \hskip 30pt \betadrhat{30001} = -\frac{9}{256}D_x\Lx1 - D_x\Rx\\
   \betadrhat{20200} &= - \frac{81}{512}D_x\Lx1 - D_x\Rx\,, \hskip 30pt
   \betadrhat{20101} = \frac{27}{128}D_x\Lx1 - D_x\Rx\,,\\
   \betadrhat{20020} &= \frac{45}{64}D_x\Lx2 + D_x\Rx\,,
   \hskip 30pt \betadrhat{20002} = -\frac{63}{256}D_x\Lx1 - D_x\Rx\,,
\label{eqn:drhatbetacoef}
\end{split}
\end{equation}
where I omit the last three indices if they all vanish.  

The coefficients of $\betadrhat{e}$ through two loops are:
\begin{equation}
\begin{split}
    \betaedrhat{0}{2} &= \frac{1}{2}\CA - \CF - \frac{1}{4}N_f
           - \frac{1}{4}D_x\,\Lx \CA-\CF\Rx\,,\qquad
    \betaedrhat{1}{1} = \frac{3}{2}\CF\,,\\[5pt]
    \betaedrhat{0}{3} &= 
          \frac{3}{8}\,\CA^2
          - \frac{5}{4}\,\CA\,\CF
          + \CF^2
          - \frac{3}{16}\,\CA\,N_f
          + \frac{3}{8}\,\CF\,N_f
          + D_x\Lx
          - \frac{1}{2}\,\CA^2
          + \frac{3}{2}\,\CA\,\CF
          - \CF^2
          + \frac{3}{32}\,\CA\,N_f\Rx\\
       &
          + D_x^2\Lx
          \frac{3}{32}\,\CA^2
          - \frac{1}{4}\,\CA\,\CF
          + \frac{9}{64}\,\CF^2\Rx\,,\\
    \betaedrhat{1}{2} &=
          - \frac{3}{8}\,\CA^2
          + \frac{7}{4}\,\CA\,\CF
          - 2\,\CF^2
          - \frac{5}{16}\,\CF\,N_f
          + D_x\Lx
          - \frac{11}{16}\,\CA\,\CF
          + \frac{1}{2}\,\CF^2\Rx\,,\\
    \betaedrhat{2}{1} &=
          - \frac{7}{64}\,\CA^2
          + \frac{61}{48}\,\CA\,\CF
          + \frac{3}{16}\,\CF^2
          + \frac{1}{16}\,\CA\,N_f
          - \frac{5}{24}\,\CF\,N_f
          + D_x\Lx
          \frac{1}{64}\,\CA^2
          - \frac{11}{96}\,\CA\,\CF\Rx\,,\\[5pt]
    \betaedrhat{0}{2100} \hskip-13pt&\hskip13pt= -\frac{9}{8}\Lx1 - D_x\Rx\,,\qquad
    \betaedrhat{0}{2010} = \frac{5}{8}\Lx2 + D_x\Rx\,,\qquad
    \betaedrhat{0}{2001} = \frac{3}{4}\Lx1 - D_x\Rx\,,\\
    \betaedrhat{0}{1200} \hskip-13pt&\hskip13pt= \frac{27}{64}\Lx1 - D_x\Rx\,,\qquad
    \betaedrhat{0}{1020} = -\frac{15}{8}\Lx2 + D_x\Rx\,,\qquad
    \betaedrhat{0}{1002} = \frac{21}{32}\Lx1 - D_x\Rx\,,\\
    \betaedrhat{0}{1101} \hskip-13pt&\hskip13pt= -\frac{9}{16}\Lx1 - D_x\Rx\,,\\
\label{eqn:drhatbetaecoef2}
\end{split}
\end{equation}
The three-loop coefficients that do not involve the quartic couplings are:
\begin{equation}
\begin{split}
    \betaedrhat{0}{4} &=
          \frac{9}{16}\,\CA^3\,\zeta_3
          - \CA^2\,\CF\,\Lx\frac{5}{16} + \frac{69}{16}\,\zeta_3\Rx
          + \CA\,\CF^2\,\Lx\frac{5}{4} + \frac{15}{2}\,\zeta_3\Rx
          - \CF^3\,\Lx\frac{5}{4} + \frac{9}{4}\,\zeta_3\Rx\\
          &
          + \CA^2\,N_f\,\Lx\frac{3}{128} - \frac{9}{32}\,\zeta_3\Rx
          - \CA\,\CF\,N_f\,\Lx\frac{15}{32} - \frac{51}{32}\,\zeta_3\Rx
          + \CF^2\,N_f\,\Lx\frac{27}{32} - \frac{33}{16}\,\zeta_3\Rx
          + N_f^2\,\Lx\frac{1}{256}\,\CA - \frac{1}{128}\,\CF\Rx\\
          &
          + D_x\LB
          - \CA^3\,\Lx\frac{7}{32} + \frac{3}{8}\,\zeta_3\Rx
          + \CA^2\,\CF\,\Lx\frac{91}{64} + \frac{135}{32}\,\zeta_3\Rx
          - \CA\,\CF^2\,\Lx\frac{13}{4}+ \frac{249}{32}\,\zeta_3\Rx
          + \CF^3\,\Lx\frac{41}{16} + \frac{27}{16}\,\zeta_3\Rx\right.\\
          &\left.
          + \CA^2\,N_f\,\Lx\frac{21}{128} + \frac{3}{64}\,\zeta_3\Rx
          - \CA\,\CF\,N_f\,\Lx\frac{37}{256}+ \frac{33}{64}\,\zeta_3\Rx
          - \CF^2\,N_f\,\Lx\frac{47}{128} - \frac{27}{32}\,\zeta_3\Rx
          - N_f^2\,\Lx\frac{1}{512}\,\CA + \frac{3}{64}\,\CF\Rx
          \RB\\
          &
          + D_x^2\LB
          + \frac{9}{64}\,\CA^3
          - \CA^2\,\CF\,\Lx\frac{35}{64} + \frac{69}{64}\,\zeta_3\Rx
          + \CA\,\CF^2\,\Lx\frac{461}{512} + \frac{147}{64}\,\zeta_3\Rx
          - \CF^3\,\Lx\frac{189}{256} + \frac{9}{32}\,\zeta_3\Rx\right.\\
          &\left.
          - \CA^2\,N_f\,\Lx\frac{29}{512} - \frac{3}{128}\,\zeta_3\Rx
          + \CA\,\CF\,N_f\,\Lx\frac{49}{512} - \frac{9}{128}\,\zeta_3\Rx
          - \CF^2\,N_f\,\Lx\frac{43}{1024} - \frac{3}{64}\,\zeta_3\Rx
          \RB\\
          &
          + D_x^3\LB
          - \CA^3\,\Lx\frac{1}{32} - \frac{3}{128}\,\zeta_3\Rx
          + \frac{33}{256}\,\CA^2\,\CF
          - \CA\,\CF^2\,\Lx\frac{189}{1024} + \frac{9}{128}\,\zeta_3\Rx
          + \CF^3\,\Lx\frac{109}{1024} - \frac{3}{64}\,\zeta_3\Rx
          \RB\,,\\
    \betaedrhat{1}{3} &=
          - \CA^3\,\Lx\frac{25}{64} - \frac{3}{4}\,\zeta_3\Rx
          + \CA^2\,\CF\,\Lx\frac{85}{32} - \frac{15}{4}\,\zeta_3\Rx
          - \CA\,\CF^2\,\Lx\frac{11}{2} - 6\,\zeta_3\Rx
          + \CF^3\,\Lx\frac{7}{2} - 3\,\zeta_3\Rx\\
          &
          + \CA^2\,N_f\,\Lx\frac{7}{32} - \frac{3}{8}\,\zeta_3\Rx
          - \CA\,\CF\,N_f\,\Lx\frac{27}{32} - \frac{9}{8}\,\zeta_3\Rx
          + \CF^2\,N_f\,\Lx\frac{13}{16} - \frac{3}{4}\,\zeta_3\Rx
          + \frac{3}{64}\,\CA\,N_f^2\\
          &
          + D_x\LB
          - \CA^3\,\Lx\frac{13}{32} + \frac{3}{4}\,\zeta_3\Rx
          + \CA^2\,\CF\Lx1  + \frac{63}{16}\,\zeta_3\Rx
          + \CA\,\CF^2\,\Lx\frac{5}{64} - \frac{105}{16}\,\zeta_3\Rx
          - \CF^3\,\Lx\frac{29}{32} - \frac{27}{8}\,\zeta_3\Rx\right.\\
          &\left.
          + \CA^2\,N_f\,\Lx\frac{1}{128} + \frac{3}{16}\,\zeta_3\Rx
          + \CA\,\CF\,N_f\,\Lx\frac{51}{128} - \frac{9}{16}\,\zeta_3\Rx
          - \CF^2\,N_f\,\Lx\frac{25}{128} - \frac{3}{8}\,\zeta_3\Rx
          \RB\\
          &
          + D_x^2\LB
          + \CA^3\,\Lx\frac{13}{128} + \frac{3}{16}\,\zeta_3\Rx
          - \CA^2\,\CF\,\Lx\frac{25}{128} + \frac{33}{32}\,\zeta_3\Rx
          - \CA\,\CF^2\,\Lx\frac{3}{128} - \frac{57}{32}\,\zeta_3\Rx
          + \CF^3\,\Lx\frac{1}{8} - \frac{15}{16}\,\zeta_3\Rx
          \RB\,,\\
    \betaedrhat{2}{2} &= 
          \CA^3\,\Lx\frac{121}{512} - \frac{45}{16}\,\zeta_3\Rx
          - \CA^2\,\CF\,\Lx\frac{167}{256} - \frac{207}{16}\,\zeta_3\Rx
          + \CA\,\CF^2\,\Lx\frac{131}{128} - 18\,\zeta_3\Rx
          - \CF^3\,\Lx\frac{85}{64} - \frac{27}{4}\,\zeta_3\Rx\\
          &
          - \CA^2\,N_f\,\Lx\frac{899}{1024} - \frac{45}{32}\,\zeta_3\Rx
          + \CA\,\CF\,N_f\,\Lx\frac{273}{128} - \frac{171}{32}\,\zeta_3\Rx
          - \CF^2\,N_f\,\Lx\frac{641}{256} - \frac{99}{16}\,\zeta_3\Rx
          - N_f^2\Lx\frac{1}{256}\,\CA - \frac{1}{16}\,\CF\Rx\\
          &
          + D_x\LB
          - \CA^3\,\Lx\frac{4355}{1024} - \frac{45}{32}\,\zeta_3\Rx
          + \CA^2\,\CF\,\Lx\frac{21071}{1024} - \frac{99}{16}\,\zeta_3\Rx
          - \CA\,\CF^2\,\Lx\frac{3381}{128} - \frac{261}{32}\,\zeta_3\Rx\right.\\
          &\left.
          + \CF^3\,\Lx\frac{13}{256} - \frac{45}{16}\,\zeta_3\Rx
          + \frac{1}{1024}\,\CA^2\,N_f
          + \frac{15}{64}\,\CA\,\CF\,N_f
          + \frac{1}{16}\,\CF^2\,N_f
          \RB\\
          &
          + D_x^2\LB
          - \frac{1}{1024}\,\CA^3
          + \frac{83}{1024}\,\CA^2\,\CF
          - \frac{33}{512}\,\CA\,\CF^2
          \RB\,,\\
    \betaedrhat{3}{1} &= 
          - \frac{3025}{4608}\,\CA^3
          + \frac{12601}{3456}\,\CA^2\,\CF
          - \frac{453}{128}\,\CA\,\CF^2
          + \frac{129}{64}\,\CF^3
          + \frac{475}{2304}\,\CA^2\,N_f
          - \CA\,\CF\,N_f\,\Lx\frac{151}{1728} + \frac{3}{4}\,\zeta_3\Rx\\
          &
          - \CF^2\,N_f\,\Lx\frac{23}{32} - \frac{3}{4}\,\zeta_3\Rx
          - \frac{5}{576}\,\CA\,N_f^2
          - \frac{35}{864}\,\CF\,N_f^2\\
          &
          + D_x\LB
          + \frac{643}{9216}\,\CA^3
          - \frac{883}{1728}\,\CA^2\,\CF
          - \frac{5}{256}\,\CA\,\CF^2
          - \frac{1}{144}\,\CA^2\,N_f
          - \frac{19}{864}\,\CA\,\CF\,N_f
          \RB\\
          &
          + D_x^2\LB
          - \frac{11}{9216}\,\CA^3
          - \frac{5}{13824}\,\CA^2\,\CF
          \RB\,,
\label{eqn:drhatbetaecoef3}
\end{split}
\end{equation}
while the three-loop coefficients that do involve the quartic
interactions are:
\begin{equation}
\begin{split}
    \betaedrhat{0}{3100} \hskip-13pt&\hskip13pt =
          - \frac{9}{64}\Lx1 - 7\,D_x + 6\,D_x^2\Rx
          + \frac{135}{128}N_f\Lx1 - D_x\Rx\,,\\
    \betaedrhat{0}{3010} \hskip-13pt&\hskip13pt =
       \frac{5}{64}\Lx8 - 18\,D_x - 11\,D_x^2\Rx
          - \frac{75}{128}\,N_f\Lx2 + D_x\Rx\,,\\
    \betaedrhat{0}{3001} \hskip-13pt&\hskip13pt =
          \frac{3}{64}\Lx2 - 19\,D_x + 17\,D_x^2\Rx
          - \frac{45}{64}\,N_f\Lx1 - D_x\Rx\,,\\
    \betaedrhat{1}{2100} \hskip-13pt&\hskip13pt =
          - \frac{51}{8}\Lx1 - D_x\Rx\,,\qquad
    \betaedrhat{1}{2010} =
          \frac{85}{24}\Lx2 + D_x\Rx\,,\qquad
    \betaedrhat{1}{2001} =
          \frac{17}{4}\Lx1 - D_x\Rx\,,\\
    \betaedrhat{2}{1100} \hskip-13pt&\hskip13pt =
          - \frac{801}{1024}\Lx1 - D_x\Rx\,,\qquad
    \betaedrhat{2}{1010} = 
          \frac{375}{256}\Lx2 + D_x\Rx\,,\qquad
    \betaedrhat{2}{1001} =
          \frac{507}{512}\Lx1 - D_x\Rx\,,\\
    \betaedrhat{0}{2200} \hskip-13pt&\hskip13pt =
          \frac{3}{1024}\Lx422 - 553\,D_x + 131\,D_x^2\Rx
          - \frac{405}{1024}\,N_f\Lx1 - D_x\Rx\,,\\
    \betaedrhat{0}{2020} \hskip-13pt&\hskip13pt =
          - \frac{5}{384}\Lx652 + 136\,D_x - 95\,D_x^2\Rx
          + \frac{225}{128}\,N_f\Lx2 + D_x\Rx\,,\\
    \betaedrhat{0}{2002} \hskip-13pt&\hskip13pt =
          \frac{1}{1536}\Lx394 - 731\,D_x + 337\,D_x^2\Rx
          - \frac{315}{512}\,N_f\Lx1 - D_x\Rx\,,\\
    \betaedrhat{0}{2110} \hskip-13pt&\hskip13pt =
          \frac{55}{32}\Lx2 - D_x - D_x^2\Rx\,,\\
    \betaedrhat{0}{2101} \hskip-13pt&\hskip13pt =
          - \frac{1}{256}\Lx622 - 773\,D_x + 151\,D_x^2\Rx
          + \frac{135}{256}\,N_f\Lx1 - D_x\Rx\,,\\
    \betaedrhat{0}{2011} \hskip-13pt&\hskip13pt =
          - \frac{205}{96}\Lx2 - D_x - D_x^2\Rx\,,\\
    \betaedrhat{1}{1200} \hskip-13pt&\hskip13pt =
          \frac{405}{128}\Lx1 - D_x\Rx\,,\qquad
    \betaedrhat{1}{1020} =
          - \frac{225}{16}\Lx2 + D_x\Rx\,,\\
    \betaedrhat{1}{1002} \hskip-13pt&\hskip13pt =
          \frac{315}{64}\Lx1 - D_x\Rx\,,\qquad
    \betaedrhat{1}{1101} =
          - \frac{135}{32}\Lx1 - D_x\Rx\,,\\
    \betaedrhat{0}{1300} \hskip-13pt&\hskip13pt =
          - \frac{27}{1024}\Lx11 - 10\,D_x - D_x^2\Rx\,,\qquad
    \betaedrhat{0}{1210} =
          - \frac{135}{256}\Lx2 - D_x - D_x^2\Rx\,,\\
    \betaedrhat{0}{1201} \hskip-13pt&\hskip13pt =
          \frac{27}{512}\Lx11 - 10\,D_x - D_x^2\Rx\,,\qquad
    \betaedrhat{0}{1120} =
          - \frac{45}{64}\Lx2 - D_x - D_x^2\Rx\,,\\
    \betaedrhat{0}{1111} \hskip-13pt&\hskip13pt =
          \frac{45}{32}\Lx2 - D_x - D_x^2\Rx\,,\qquad
    \betaedrhat{0}{1102} =
          \frac{9}{256}\Lx14 - 25\,D_x + 11\,D_x^2\Rx\,,\\
    \betaedrhat{0}{1030} \hskip-13pt&\hskip13pt =
          \frac{5}{4}\Lx16 + 10\,D_x + D_x^2\Rx\,,\qquad
    \betaedrhat{0}{1021} =
          \frac{105}{64}\Lx2 - D_x - D_x^2\Rx\,,\\
    \betaedrhat{0}{1012} \hskip-13pt&\hskip13pt =
          - \frac{105}{64}\Lx2 - D_x - D_x^2\Rx\,,\qquad
    \betaedrhat{0}{1003} =
          - \frac{7}{256}\Lx14 - 25\,D_x + 11\,D_x^2\Rx\,,
\label{eqn:drhatbetaecoef4}
\end{split}
\end{equation}

A consistent description of $\betadrhat{}$ and $\betaedrhat{}{}$
through three loops only requires knowledge of the
$\betadrhat{\eta_i}$'s through one loop.  These coefficients are:
\begin{equation}
\begin{split}
   \betadrhat{\eta_1,\,20000} &= - \frac{3}{8}\,,\qquad
   \betadrhat{\eta_1,\,02000} =  \frac{1}{3}N_f\,,\qquad
   \betadrhat{\eta_1,\,10100} =  \frac{9}{2}\,,\qquad
   \betadrhat{\eta_1,\,01100} = - \frac{1}{2}N_f\,,\\
   \betadrhat{\eta_1,\,00200} &= - \frac{11}{8} - \frac{1}{8}\,D_x\,,\qquad
   \betadrhat{\eta_1,\,00110} = - 2 - D_x\,,\qquad
   \betadrhat{\eta_1,\,00101} = \frac{7}{2} - \frac{1}{2}\,D_x\,,\\
   \betadrhat{\eta_2,\,20000} &= - \frac{9}{16}\,,\qquad
   \betadrhat{\eta_2,\,02000} = \frac{1}{24}N_f\,,\qquad
   \betadrhat{\eta_2,\,10010} = \frac{9}{2}\,,\qquad
   \betadrhat{\eta_2,\,01010} = - \frac{1}{2}N_f\,,\\
   \betadrhat{\eta_2,\,00200} &= \frac{3}{16}\Lx1 - D_x\Rx\,,\qquad
   \betadrhat{\eta_2,\,00110} = \frac{1}{2}\Lx1 - D_x\Rx\,,\qquad
   \betadrhat{\eta_2,\,00101} = - \frac{1}{2}\Lx1 - D_x\Rx\,,\\
   \betadrhat{\eta_2,\,00020} &= - \frac{32}{3} - \frac{4}{3}\,D_x\,,\qquad
   \betadrhat{\eta_2,\,00011} = - \frac{7}{6}\Lx1 - D_x\Rx\,,\qquad
   \betadrhat{\eta_2,\,00002} = \frac{7}{12}\Lx1 - D_x\Rx\,,\\
   \betadrhat{\eta_3,\,10001} &= \frac{9}{2}\,,\qquad
   \betadrhat{\eta_3,\,01001} = - \frac{1}{2}N_f\,,\qquad
   \betadrhat{\eta_3,\,00110} = 2 + D_x\,,\qquad
   \betadrhat{\eta_3,\,00101} = \frac{5}{2} - D_x\,,\\
   \betadrhat{\eta_3,\,00020} &= \frac{5}{3}\Lx2 + D_x\Rx\,,\qquad
   \betadrhat{\eta_3,\,00011} = - \frac{10}{3}\Lx2 + D_x\Rx\,,\qquad
   \betadrhat{\eta_3,\,00002} = - \frac{7}{6} + \frac{11}{12}\,D_x\,,\qquad
\label{eqn:drhatbetaetacoef}
\end{split}
\end{equation}


\begin{thebibliography}{57}
\expandafter\ifx\csname natexlab\endcsname\relax\def\natexlab#1{#1}\fi
\expandafter\ifx\csname bibnamefont\endcsname\relax
  \def\bibnamefont#1{#1}\fi
\expandafter\ifx\csname bibfnamefont\endcsname\relax
  \def\bibfnamefont#1{#1}\fi
\expandafter\ifx\csname citenamefont\endcsname\relax
  \def\citenamefont#1{#1}\fi
\expandafter\ifx\csname url\endcsname\relax
  \def\url#1{\texttt{#1}}\fi
\expandafter\ifx\csname urlprefix\endcsname\relax\def\urlprefix{URL }\fi
\providecommand{\bibinfo}[2]{#2}
\providecommand{\eprint}[2][]{\url{#2}}

\bibitem[{\citenamefont{Bern and Kosower}(1992)}]{Bern:1992aq}
\bibinfo{author}{\bibfnamefont{Z.}~\bibnamefont{Bern}} \bibnamefont{and}
  \bibinfo{author}{\bibfnamefont{D.~A.} \bibnamefont{Kosower}},
  \bibinfo{journal}{Nucl. Phys.} \textbf{\bibinfo{volume}{B379}},
  \bibinfo{pages}{451} (\bibinfo{year}{1992}).

\bibitem[{\citenamefont{Bern et~al.}(2002)\citenamefont{Bern, De~Freitas,
  Dixon, and Wong}}]{Bern:2002zk}
\bibinfo{author}{\bibfnamefont{Z.}~\bibnamefont{Bern}},
  \bibinfo{author}{\bibfnamefont{A.}~\bibnamefont{De~Freitas}},
  \bibinfo{author}{\bibfnamefont{L.~J.} \bibnamefont{Dixon}}, \bibnamefont{and}
  \bibinfo{author}{\bibfnamefont{H.~L.} \bibnamefont{Wong}},
  \bibinfo{journal}{Phys. Rev.} \textbf{\bibinfo{volume}{D66}},
  \bibinfo{pages}{085002} (\bibinfo{year}{2002}), \eprint{hep-ph/0202271}.

\bibitem[{\citenamefont{Kilgore}(2011)}]{Kilgore:2011ta}
\bibinfo{author}{\bibfnamefont{W.~B.} \bibnamefont{Kilgore}},
  \bibinfo{journal}{Phys.Rev.} \textbf{\bibinfo{volume}{D83}},
  \bibinfo{pages}{114005} (\bibinfo{year}{2011}), \eprint{1102.5353}.

\bibitem[{\citenamefont{'t~Hooft and Veltman}(1972)}]{'tHooft:1972fi}
\bibinfo{author}{\bibfnamefont{G.}~\bibnamefont{'t~Hooft}} \bibnamefont{and}
  \bibinfo{author}{\bibfnamefont{M.~J.~G.} \bibnamefont{Veltman}},
  \bibinfo{journal}{Nucl. Phys.} \textbf{\bibinfo{volume}{B44}},
  \bibinfo{pages}{189} (\bibinfo{year}{1972}).

\bibitem[{\citenamefont{Kunszt et~al.}(1994)\citenamefont{Kunszt, Signer, and
  Trocsanyi}}]{Kunszt:1993sd}
\bibinfo{author}{\bibfnamefont{Z.}~\bibnamefont{Kunszt}},
  \bibinfo{author}{\bibfnamefont{A.}~\bibnamefont{Signer}}, \bibnamefont{and}
  \bibinfo{author}{\bibfnamefont{Z.}~\bibnamefont{Trocsanyi}},
  \bibinfo{journal}{Nucl.Phys.} \textbf{\bibinfo{volume}{B411}},
  \bibinfo{pages}{397} (\bibinfo{year}{1994}), \eprint{hep-ph/9305239}.

\bibitem[{\citenamefont{Boughezal et~al.}(2011)\citenamefont{Boughezal,
  Melnikov, and Petriello}}]{Boughezal:2011br}
\bibinfo{author}{\bibfnamefont{R.}~\bibnamefont{Boughezal}},
  \bibinfo{author}{\bibfnamefont{K.}~\bibnamefont{Melnikov}}, \bibnamefont{and}
  \bibinfo{author}{\bibfnamefont{F.}~\bibnamefont{Petriello}},
  \bibinfo{journal}{Phys.Rev.} \textbf{\bibinfo{volume}{D84}},
  \bibinfo{pages}{034044} (\bibinfo{year}{2011}), \eprint{1106.5520}.

\bibitem[{\citenamefont{Siegel}(1979)}]{Siegel:1979wq}
\bibinfo{author}{\bibfnamefont{W.}~\bibnamefont{Siegel}},
  \bibinfo{journal}{Phys. Lett.} \textbf{\bibinfo{volume}{B84}},
  \bibinfo{pages}{193} (\bibinfo{year}{1979}).

\bibitem[{\citenamefont{Collins}(1984)}]{Collins:Renorm}
\bibinfo{author}{\bibfnamefont{J.}~\bibnamefont{Collins}},
  \emph{\bibinfo{title}{Renormalization}} (\bibinfo{publisher}{Cambridge
  University Press}, \bibinfo{address}{Cambridge, England},
  \bibinfo{year}{1984}).

\bibitem[{\citenamefont{Jack et~al.}(1994{\natexlab{a}})\citenamefont{Jack,
  Jones, and Roberts}}]{Jack:1993ws}
\bibinfo{author}{\bibfnamefont{I.}~\bibnamefont{Jack}},
  \bibinfo{author}{\bibfnamefont{D.~R.~T.} \bibnamefont{Jones}},
  \bibnamefont{and} \bibinfo{author}{\bibfnamefont{K.~L.}
  \bibnamefont{Roberts}}, \bibinfo{journal}{Z. Phys.}
  \textbf{\bibinfo{volume}{C62}}, \bibinfo{pages}{161}
  (\bibinfo{year}{1994}{\natexlab{a}}), \eprint{hep-ph/9310301}.

\bibitem[{\citenamefont{Siegel}(1980)}]{Siegel:1980qs}
\bibinfo{author}{\bibfnamefont{W.}~\bibnamefont{Siegel}},
  \bibinfo{journal}{Phys.Lett.} \textbf{\bibinfo{volume}{B94}},
  \bibinfo{pages}{37} (\bibinfo{year}{1980}).

\bibitem[{\citenamefont{St{\"o}ckinger}(2005)}]{Stockinger:2005gx}
\bibinfo{author}{\bibfnamefont{D.}~\bibnamefont{St{\"o}ckinger}},
  \bibinfo{journal}{JHEP} \textbf{\bibinfo{volume}{0503}}, \bibinfo{pages}{076}
  (\bibinfo{year}{2005}), \eprint{hep-ph/0503129}.

\bibitem[{\citenamefont{Signer and St{\"o}ckinger}(2009)}]{Signer:2008va}
\bibinfo{author}{\bibfnamefont{A.}~\bibnamefont{Signer}} \bibnamefont{and}
  \bibinfo{author}{\bibfnamefont{D.}~\bibnamefont{St{\"o}ckinger}},
  \bibinfo{journal}{Nucl. Phys.} \textbf{\bibinfo{volume}{B808}},
  \bibinfo{pages}{88} (\bibinfo{year}{2009}), \eprint{0807.4424}.

\bibitem[{\citenamefont{Kosower}(1991)}]{Kosower:1990ax}
\bibinfo{author}{\bibfnamefont{D.~A.} \bibnamefont{Kosower}},
  \bibinfo{journal}{Phys.Lett.} \textbf{\bibinfo{volume}{B254}},
  \bibinfo{pages}{439} (\bibinfo{year}{1991}).

\bibitem[{\citenamefont{Catani}(1998)}]{Catani:1998bh}
\bibinfo{author}{\bibfnamefont{S.}~\bibnamefont{Catani}},
  \bibinfo{journal}{Phys. Lett.} \textbf{\bibinfo{volume}{B427}},
  \bibinfo{pages}{161} (\bibinfo{year}{1998}),
  \eprint[http://arXiv.org/abs]{hep-ph/9802439}.

\bibitem[{\citenamefont{Sterman and Tejeda-Yeomans}(2003)}]{Sterman:2002qn}
\bibinfo{author}{\bibfnamefont{G.}~\bibnamefont{Sterman}} \bibnamefont{and}
  \bibinfo{author}{\bibfnamefont{M.~E.} \bibnamefont{Tejeda-Yeomans}},
  \bibinfo{journal}{Phys. Lett.} \textbf{\bibinfo{volume}{B552}},
  \bibinfo{pages}{48} (\bibinfo{year}{2003}),
  \eprint[http://arXiv.org/abs]{hep-ph/0210130}.

\bibitem[{\citenamefont{Aybat et~al.}(2006{\natexlab{a}})\citenamefont{Aybat,
  Dixon, and Sterman}}]{Aybat:2006wq}
\bibinfo{author}{\bibfnamefont{S.}~\bibnamefont{Aybat}},
  \bibinfo{author}{\bibfnamefont{L.~J.} \bibnamefont{Dixon}}, \bibnamefont{and}
  \bibinfo{author}{\bibfnamefont{G.~F.} \bibnamefont{Sterman}},
  \bibinfo{journal}{Phys.Rev.Lett.} \textbf{\bibinfo{volume}{97}},
  \bibinfo{pages}{072001} (\bibinfo{year}{2006}{\natexlab{a}}),
  \eprint{hep-ph/0606254}.

\bibitem[{\citenamefont{Aybat et~al.}(2006{\natexlab{b}})\citenamefont{Aybat,
  Dixon, and Sterman}}]{Aybat:2006mz}
\bibinfo{author}{\bibfnamefont{S.}~\bibnamefont{Aybat}},
  \bibinfo{author}{\bibfnamefont{L.~J.} \bibnamefont{Dixon}}, \bibnamefont{and}
  \bibinfo{author}{\bibfnamefont{G.~F.} \bibnamefont{Sterman}},
  \bibinfo{journal}{Phys.Rev.} \textbf{\bibinfo{volume}{D74}},
  \bibinfo{pages}{074004} (\bibinfo{year}{2006}{\natexlab{b}}),
  \eprint{hep-ph/0607309}.

\bibitem[{\citenamefont{Becher and
  Neubert}(2009{\natexlab{a}})}]{Becher:2009cu}
\bibinfo{author}{\bibfnamefont{T.}~\bibnamefont{Becher}} \bibnamefont{and}
  \bibinfo{author}{\bibfnamefont{M.}~\bibnamefont{Neubert}},
  \bibinfo{journal}{Phys.Rev.Lett.} \textbf{\bibinfo{volume}{102}},
  \bibinfo{pages}{162001} (\bibinfo{year}{2009}{\natexlab{a}}),
  \eprint{0901.0722}.

\bibitem[{\citenamefont{Becher and
  Neubert}(2009{\natexlab{b}})}]{Becher:2009qa}
\bibinfo{author}{\bibfnamefont{T.}~\bibnamefont{Becher}} \bibnamefont{and}
  \bibinfo{author}{\bibfnamefont{M.}~\bibnamefont{Neubert}},
  \bibinfo{journal}{JHEP} \textbf{\bibinfo{volume}{0906}}, \bibinfo{pages}{081}
  (\bibinfo{year}{2009}{\natexlab{b}}), \eprint{0903.1126}.

\bibitem[{\citenamefont{Gardi and Magnea}(2009{\natexlab{a}})}]{Gardi:2009qi}
\bibinfo{author}{\bibfnamefont{E.}~\bibnamefont{Gardi}} \bibnamefont{and}
  \bibinfo{author}{\bibfnamefont{L.}~\bibnamefont{Magnea}},
  \bibinfo{journal}{JHEP} \textbf{\bibinfo{volume}{0903}}, \bibinfo{pages}{079}
  (\bibinfo{year}{2009}{\natexlab{a}}), \eprint{0901.1091}.

\bibitem[{\citenamefont{Becher and
  Neubert}(2009{\natexlab{c}})}]{Becher:2009kw}
\bibinfo{author}{\bibfnamefont{T.}~\bibnamefont{Becher}} \bibnamefont{and}
  \bibinfo{author}{\bibfnamefont{M.}~\bibnamefont{Neubert}},
  \bibinfo{journal}{Phys.Rev.} \textbf{\bibinfo{volume}{D79}},
  \bibinfo{pages}{125004} (\bibinfo{year}{2009}{\natexlab{c}}),
  \eprint{0904.1021}.

\bibitem[{\citenamefont{Gardi and Magnea}(2009{\natexlab{b}})}]{Gardi:2009zv}
\bibinfo{author}{\bibfnamefont{E.}~\bibnamefont{Gardi}} \bibnamefont{and}
  \bibinfo{author}{\bibfnamefont{L.}~\bibnamefont{Magnea}},
  \bibinfo{journal}{Nuovo Cim.} \textbf{\bibinfo{volume}{C32N5-6}},
  \bibinfo{pages}{137} (\bibinfo{year}{2009}{\natexlab{b}}),
  \eprint{0908.3273}.

\bibitem[{\citenamefont{Dixon et~al.}(2010)\citenamefont{Dixon, Gardi, and
  Magnea}}]{Dixon:2009ur}
\bibinfo{author}{\bibfnamefont{L.~J.} \bibnamefont{Dixon}},
  \bibinfo{author}{\bibfnamefont{E.}~\bibnamefont{Gardi}}, \bibnamefont{and}
  \bibinfo{author}{\bibfnamefont{L.}~\bibnamefont{Magnea}},
  \bibinfo{journal}{JHEP} \textbf{\bibinfo{volume}{1002}}, \bibinfo{pages}{081}
  (\bibinfo{year}{2010}), \eprint{0910.3653}.

\bibitem[{\citenamefont{Mitov et~al.}(2009)\citenamefont{Mitov, Sterman, and
  Sung}}]{Mitov:2009sv}
\bibinfo{author}{\bibfnamefont{A.}~\bibnamefont{Mitov}},
  \bibinfo{author}{\bibfnamefont{G.~F.} \bibnamefont{Sterman}},
  \bibnamefont{and} \bibinfo{author}{\bibfnamefont{I.}~\bibnamefont{Sung}},
  \bibinfo{journal}{Phys.Rev.} \textbf{\bibinfo{volume}{D79}},
  \bibinfo{pages}{094015} (\bibinfo{year}{2009}), \eprint{0903.3241}.

\bibitem[{\citenamefont{Mitov et~al.}(2010)\citenamefont{Mitov, Sterman, and
  Sung}}]{Mitov:2010xw}
\bibinfo{author}{\bibfnamefont{A.}~\bibnamefont{Mitov}},
  \bibinfo{author}{\bibfnamefont{G.~F.} \bibnamefont{Sterman}},
  \bibnamefont{and} \bibinfo{author}{\bibfnamefont{I.}~\bibnamefont{Sung}},
  \bibinfo{journal}{Phys.Rev.} \textbf{\bibinfo{volume}{D82}},
  \bibinfo{pages}{034020} (\bibinfo{year}{2010}), \eprint{1005.4646}.

\bibitem[{\citenamefont{Catani and Seymour}(1996)}]{Catani:1996jh}
\bibinfo{author}{\bibfnamefont{S.}~\bibnamefont{Catani}} \bibnamefont{and}
  \bibinfo{author}{\bibfnamefont{M.~H.} \bibnamefont{Seymour}},
  \bibinfo{journal}{Phys. Lett.} \textbf{\bibinfo{volume}{B378}},
  \bibinfo{pages}{287} (\bibinfo{year}{1996}),
  \eprint[http://arXiv.org/abs]{hep-ph/9602277}.

\bibitem[{\citenamefont{Catani and Seymour}(1997)}]{Catani:1997vz}
\bibinfo{author}{\bibfnamefont{S.}~\bibnamefont{Catani}} \bibnamefont{and}
  \bibinfo{author}{\bibfnamefont{M.~H.} \bibnamefont{Seymour}},
  \bibinfo{journal}{Nucl. Phys.} \textbf{\bibinfo{volume}{B485}},
  \bibinfo{pages}{291} (\bibinfo{year}{1997}),
  \eprint[http://arXiv.org/abs]{hep-ph/9605323}.

\bibitem[{\citenamefont{Gonsalves}(1983)}]{Gonsalves:1983nq}
\bibinfo{author}{\bibfnamefont{R.~J.} \bibnamefont{Gonsalves}},
  \bibinfo{journal}{Phys. Rev.} \textbf{\bibinfo{volume}{D28}},
  \bibinfo{pages}{1542} (\bibinfo{year}{1983}).

\bibitem[{\citenamefont{Kramer and Lampe}(1987)}]{Kramer:1986sg}
\bibinfo{author}{\bibfnamefont{G.}~\bibnamefont{Kramer}} \bibnamefont{and}
  \bibinfo{author}{\bibfnamefont{B.}~\bibnamefont{Lampe}},
  \bibinfo{journal}{Z.Phys.} \textbf{\bibinfo{volume}{C34}},
  \bibinfo{pages}{497} (\bibinfo{year}{1987}).

\bibitem[{\citenamefont{Matsuura and van Neerven}(1988)}]{Matsuura:1987wt}
\bibinfo{author}{\bibfnamefont{T.}~\bibnamefont{Matsuura}} \bibnamefont{and}
  \bibinfo{author}{\bibfnamefont{W.~L.} \bibnamefont{van Neerven}},
  \bibinfo{journal}{Z. Phys.} \textbf{\bibinfo{volume}{C38}},
  \bibinfo{pages}{623} (\bibinfo{year}{1988}).

\bibitem[{\citenamefont{Matsuura et~al.}(1989)\citenamefont{Matsuura, van~der
  Marck, and van Neerven}}]{Matsuura:1989sm}
\bibinfo{author}{\bibfnamefont{T.}~\bibnamefont{Matsuura}},
  \bibinfo{author}{\bibfnamefont{S.~C.} \bibnamefont{van~der Marck}},
  \bibnamefont{and} \bibinfo{author}{\bibfnamefont{W.~L.} \bibnamefont{van
  Neerven}}, \bibinfo{journal}{Nucl. Phys.} \textbf{\bibinfo{volume}{B319}},
  \bibinfo{pages}{570} (\bibinfo{year}{1989}).

\bibitem[{\citenamefont{Harlander}(2000)}]{Harlander:2000mg}
\bibinfo{author}{\bibfnamefont{R.~V.} \bibnamefont{Harlander}},
  \bibinfo{journal}{Phys. Lett.} \textbf{\bibinfo{volume}{B492}},
  \bibinfo{pages}{74} (\bibinfo{year}{2000}),
  \eprint[http://arXiv.org/abs]{hep-ph/0007289}.

\bibitem[{\citenamefont{Moch et~al.}(2005{\natexlab{a}})\citenamefont{Moch,
  Vermaseren, and Vogt}}]{Moch:2005id}
\bibinfo{author}{\bibfnamefont{S.}~\bibnamefont{Moch}},
  \bibinfo{author}{\bibfnamefont{J.}~\bibnamefont{Vermaseren}},
  \bibnamefont{and} \bibinfo{author}{\bibfnamefont{A.}~\bibnamefont{Vogt}},
  \bibinfo{journal}{JHEP} \textbf{\bibinfo{volume}{0508}}, \bibinfo{pages}{049}
  (\bibinfo{year}{2005}{\natexlab{a}}), \eprint{hep-ph/0507039}.

\bibitem[{\citenamefont{Moch et~al.}(2005{\natexlab{b}})\citenamefont{Moch,
  Vermaseren, and Vogt}}]{Moch:2005tm}
\bibinfo{author}{\bibfnamefont{S.}~\bibnamefont{Moch}},
  \bibinfo{author}{\bibfnamefont{J.}~\bibnamefont{Vermaseren}},
  \bibnamefont{and} \bibinfo{author}{\bibfnamefont{A.}~\bibnamefont{Vogt}},
  \bibinfo{journal}{Phys.Lett.} \textbf{\bibinfo{volume}{B625}},
  \bibinfo{pages}{245} (\bibinfo{year}{2005}{\natexlab{b}}),
  \eprint{hep-ph/0508055}.

\bibitem[{\citenamefont{Bern et~al.}(1998)\citenamefont{Bern, Del~Duca, and
  Schmidt}}]{Bern:1998sc}
\bibinfo{author}{\bibfnamefont{Z.}~\bibnamefont{Bern}},
  \bibinfo{author}{\bibfnamefont{V.}~\bibnamefont{Del~Duca}}, \bibnamefont{and}
  \bibinfo{author}{\bibfnamefont{C.~R.} \bibnamefont{Schmidt}},
  \bibinfo{journal}{Phys. Lett.} \textbf{\bibinfo{volume}{B445}},
  \bibinfo{pages}{168} (\bibinfo{year}{1998}), \eprint{hep-ph/9810409}.

\bibitem[{\citenamefont{Bern et~al.}(1999)\citenamefont{Bern, Duca, Kilgore,
  and Schmidt}}]{Bern:1999ry}
\bibinfo{author}{\bibfnamefont{Z.}~\bibnamefont{Bern}},
  \bibinfo{author}{\bibfnamefont{V.~D.} \bibnamefont{Duca}},
  \bibinfo{author}{\bibfnamefont{W.~B.} \bibnamefont{Kilgore}},
  \bibnamefont{and} \bibinfo{author}{\bibfnamefont{C.~R.}
  \bibnamefont{Schmidt}}, \bibinfo{journal}{Phys. Rev.}
  \textbf{\bibinfo{volume}{D60}}, \bibinfo{pages}{116001}
  (\bibinfo{year}{1999}), \eprint[http://arXiv.org/abs]{hep-ph/9903516}.

\bibitem[{\citenamefont{Kosower and Uwer}(1999)}]{Kosower:1999rx}
\bibinfo{author}{\bibfnamefont{D.~A.} \bibnamefont{Kosower}} \bibnamefont{and}
  \bibinfo{author}{\bibfnamefont{P.}~\bibnamefont{Uwer}},
  \bibinfo{journal}{Nucl. Phys.} \textbf{\bibinfo{volume}{B563}},
  \bibinfo{pages}{477} (\bibinfo{year}{1999}),
  \eprint[http://arXiv.org/abs]{hep-ph/9903515}.

\bibitem[{\citenamefont{Jack et~al.}(1994{\natexlab{b}})\citenamefont{Jack,
  Jones, and Roberts}}]{Jack:1994bn}
\bibinfo{author}{\bibfnamefont{I.}~\bibnamefont{Jack}},
  \bibinfo{author}{\bibfnamefont{D.~R.~T.} \bibnamefont{Jones}},
  \bibnamefont{and} \bibinfo{author}{\bibfnamefont{K.~L.}
  \bibnamefont{Roberts}}, \bibinfo{journal}{Z. Phys.}
  \textbf{\bibinfo{volume}{C63}}, \bibinfo{pages}{151}
  (\bibinfo{year}{1994}{\natexlab{b}}), \eprint{hep-ph/9401349}.

\bibitem[{\citenamefont{Harlander et~al.}(2006)\citenamefont{Harlander, Kant,
  Mihaila, and Steinhauser}}]{Harlander:2006rj}
\bibinfo{author}{\bibfnamefont{R.}~\bibnamefont{Harlander}},
  \bibinfo{author}{\bibfnamefont{P.}~\bibnamefont{Kant}},
  \bibinfo{author}{\bibfnamefont{L.}~\bibnamefont{Mihaila}}, \bibnamefont{and}
  \bibinfo{author}{\bibfnamefont{M.}~\bibnamefont{Steinhauser}},
  \bibinfo{journal}{JHEP} \textbf{\bibinfo{volume}{09}}, \bibinfo{pages}{053}
  (\bibinfo{year}{2006}), \eprint{hep-ph/0607240}.

\bibitem[{\citenamefont{Capper et~al.}(1980)\citenamefont{Capper, Jones, and
  van Nieuwenhuizen}}]{Capper:1980ns}
\bibinfo{author}{\bibfnamefont{D.~M.} \bibnamefont{Capper}},
  \bibinfo{author}{\bibfnamefont{D.~R.~T.} \bibnamefont{Jones}},
  \bibnamefont{and} \bibinfo{author}{\bibfnamefont{P.}~\bibnamefont{van
  Nieuwenhuizen}}, \bibinfo{journal}{Nucl. Phys.}
  \textbf{\bibinfo{volume}{B167}}, \bibinfo{pages}{479} (\bibinfo{year}{1980}).

\bibitem[{\citenamefont{Anastasiou et~al.}(2001)\citenamefont{Anastasiou,
  Glover, Oleari, and Tejeda-Yeomans}}]{Anastasiou:2001sv}
\bibinfo{author}{\bibfnamefont{C.}~\bibnamefont{Anastasiou}},
  \bibinfo{author}{\bibfnamefont{E.~W.~N.} \bibnamefont{Glover}},
  \bibinfo{author}{\bibfnamefont{C.}~\bibnamefont{Oleari}}, \bibnamefont{and}
  \bibinfo{author}{\bibfnamefont{M.~E.} \bibnamefont{Tejeda-Yeomans}},
  \bibinfo{journal}{Nucl. Phys.} \textbf{\bibinfo{volume}{B605}},
  \bibinfo{pages}{486} (\bibinfo{year}{2001}),
  \eprint[http://arXiv.org/abs]{hep-ph/0101304}.

\bibitem[{\citenamefont{Anastasiou et~al.}(2002)\citenamefont{Anastasiou,
  Glover, and Tejeda-Yeomans}}]{Anastasiou:2002zn}
\bibinfo{author}{\bibfnamefont{C.}~\bibnamefont{Anastasiou}},
  \bibinfo{author}{\bibfnamefont{E.}~\bibnamefont{Glover}}, \bibnamefont{and}
  \bibinfo{author}{\bibfnamefont{M.}~\bibnamefont{Tejeda-Yeomans}},
  \bibinfo{journal}{Nucl.Phys.} \textbf{\bibinfo{volume}{B629}},
  \bibinfo{pages}{255} (\bibinfo{year}{2002}), \eprint{hep-ph/0201274}.

\bibitem[{\citenamefont{Glover and Tejeda-Yeomans}(2003)}]{Glover:2003cm}
\bibinfo{author}{\bibfnamefont{E.~N.} \bibnamefont{Glover}} \bibnamefont{and}
  \bibinfo{author}{\bibfnamefont{M.}~\bibnamefont{Tejeda-Yeomans}},
  \bibinfo{journal}{JHEP} \textbf{\bibinfo{volume}{0306}}, \bibinfo{pages}{033}
  (\bibinfo{year}{2003}), \eprint{hep-ph/0304169}.

\bibitem[{\citenamefont{Chetyrkin et~al.}(1997)\citenamefont{Chetyrkin, Kniehl,
  and Steinhauser}}]{Chetyrkin:1997iv}
\bibinfo{author}{\bibfnamefont{K.~G.} \bibnamefont{Chetyrkin}},
  \bibinfo{author}{\bibfnamefont{B.~A.} \bibnamefont{Kniehl}},
  \bibnamefont{and}
  \bibinfo{author}{\bibfnamefont{M.}~\bibnamefont{Steinhauser}},
  \bibinfo{journal}{Phys. Rev. Lett.} \textbf{\bibinfo{volume}{79}},
  \bibinfo{pages}{353} (\bibinfo{year}{1997}),
  \eprint[http://arXiv.org/abs]{hep-ph/9705240}.

\bibitem[{\citenamefont{Chetyrkin et~al.}(1998)\citenamefont{Chetyrkin, Kniehl,
  and Steinhauser}}]{Chetyrkin:1998un}
\bibinfo{author}{\bibfnamefont{K.~G.} \bibnamefont{Chetyrkin}},
  \bibinfo{author}{\bibfnamefont{B.~A.} \bibnamefont{Kniehl}},
  \bibnamefont{and}
  \bibinfo{author}{\bibfnamefont{M.}~\bibnamefont{Steinhauser}},
  \bibinfo{journal}{Nucl. Phys.} \textbf{\bibinfo{volume}{B510}},
  \bibinfo{pages}{61} (\bibinfo{year}{1998}),
  \eprint[http://arXiv.org/abs]{hep-ph/9708255}.

\bibitem[{\citenamefont{Nogueira}(1993)}]{Nogueira:1993ex}
\bibinfo{author}{\bibfnamefont{P.}~\bibnamefont{Nogueira}},
  \bibinfo{journal}{J. Comput. Phys.} \textbf{\bibinfo{volume}{105}},
  \bibinfo{pages}{279} (\bibinfo{year}{1993}).

\bibitem[{\citenamefont{Vermaseren}(2000)}]{Vermaseren:2000nd}
\bibinfo{author}{\bibfnamefont{J.~A.~M.} \bibnamefont{Vermaseren}}
  (\bibinfo{year}{2000}), \bibinfo{note}{{Report} {No.} {NIKHEF}-00-0032},
  \eprint[http://arXiv.org/abs]{math-ph/0010025}.

\bibitem[{\citenamefont{Davydychev et~al.}(1998)\citenamefont{Davydychev,
  Osland, and Tarasov}}]{Davydychev:1997vh}
\bibinfo{author}{\bibfnamefont{A.~I.} \bibnamefont{Davydychev}},
  \bibinfo{author}{\bibfnamefont{P.}~\bibnamefont{Osland}}, \bibnamefont{and}
  \bibinfo{author}{\bibfnamefont{O.}~\bibnamefont{Tarasov}},
  \bibinfo{journal}{Phys.Rev.} \textbf{\bibinfo{volume}{D58}},
  \bibinfo{pages}{036007} (\bibinfo{year}{1998}), \eprint{hep-ph/9801380}.

\bibitem[{\citenamefont{von Manteuffel and
  Studerus}(2012)}]{vonManteuffel:2012yz}
\bibinfo{author}{\bibfnamefont{A.}~\bibnamefont{von Manteuffel}}
  \bibnamefont{and} \bibinfo{author}{\bibfnamefont{C.}~\bibnamefont{Studerus}}
  (\bibinfo{year}{2012}), \eprint{1201.4330}.

\bibitem[{\citenamefont{Studerus}(2010)}]{Studerus:2009ye}
\bibinfo{author}{\bibfnamefont{C.}~\bibnamefont{Studerus}},
  \bibinfo{journal}{Comput. Phys. Commun.} \textbf{\bibinfo{volume}{181}},
  \bibinfo{pages}{1293} (\bibinfo{year}{2010}), \eprint{0912.2546}.

\bibitem[{\citenamefont{Chetyrkin et~al.}(1980)\citenamefont{Chetyrkin, Kataev,
  and Tkachov}}]{Chetyrkin:1980pr}
\bibinfo{author}{\bibfnamefont{K.~G.} \bibnamefont{Chetyrkin}},
  \bibinfo{author}{\bibfnamefont{A.~L.} \bibnamefont{Kataev}},
  \bibnamefont{and} \bibinfo{author}{\bibfnamefont{F.~V.}
  \bibnamefont{Tkachov}}, \bibinfo{journal}{Nucl. Phys.}
  \textbf{\bibinfo{volume}{B174}}, \bibinfo{pages}{345} (\bibinfo{year}{1980}).

\bibitem[{\citenamefont{Kazakov}(1984)}]{Kazakov:1983ns}
\bibinfo{author}{\bibfnamefont{D.~I.} \bibnamefont{Kazakov}},
  \bibinfo{journal}{Theor. Math. Phys.} \textbf{\bibinfo{volume}{58}},
  \bibinfo{pages}{223} (\bibinfo{year}{1984}).

\bibitem[{\citenamefont{Gehrmann et~al.}(2005)\citenamefont{Gehrmann, Huber,
  and Maitre}}]{Gehrmann:2005pd}
\bibinfo{author}{\bibfnamefont{T.}~\bibnamefont{Gehrmann}},
  \bibinfo{author}{\bibfnamefont{T.}~\bibnamefont{Huber}}, \bibnamefont{and}
  \bibinfo{author}{\bibfnamefont{D.}~\bibnamefont{Maitre}},
  \bibinfo{journal}{Phys. Lett.} \textbf{\bibinfo{volume}{B622}},
  \bibinfo{pages}{295} (\bibinfo{year}{2005}), \eprint{hep-ph/0507061}.

\bibitem[{\citenamefont{Smirnov}(1999)}]{Smirnov:1999gc}
\bibinfo{author}{\bibfnamefont{V.~A.} \bibnamefont{Smirnov}},
  \bibinfo{journal}{Phys. Lett.} \textbf{\bibinfo{volume}{B460}},
  \bibinfo{pages}{397} (\bibinfo{year}{1999}),
  \eprint[http://arXiv.org/abs]{hep-ph/9905323}.

\bibitem[{\citenamefont{Anastasiou
  et~al.}(2000{\natexlab{a}})\citenamefont{Anastasiou, Tausk, and
  Tejeda-Yeomans}}]{Anastasiou:2000kp}
\bibinfo{author}{\bibfnamefont{C.}~\bibnamefont{Anastasiou}},
  \bibinfo{author}{\bibfnamefont{J.~B.} \bibnamefont{Tausk}}, \bibnamefont{and}
  \bibinfo{author}{\bibfnamefont{M.~E.} \bibnamefont{Tejeda-Yeomans}},
  \bibinfo{journal}{Nucl. Phys. Proc. Suppl.} \textbf{\bibinfo{volume}{89}},
  \bibinfo{pages}{262} (\bibinfo{year}{2000}{\natexlab{a}}),
  \eprint{hep-ph/0005328}.

\bibitem[{\citenamefont{Tausk}(1999)}]{Tausk:1999vh}
\bibinfo{author}{\bibfnamefont{J.~B.} \bibnamefont{Tausk}},
  \bibinfo{journal}{Phys. Lett.} \textbf{\bibinfo{volume}{B469}},
  \bibinfo{pages}{225} (\bibinfo{year}{1999}),
  \eprint[http://arXiv.org/abs]{hep-ph/9909506}.

\bibitem[{\citenamefont{Anastasiou
  et~al.}(2000{\natexlab{b}})\citenamefont{Anastasiou, Gehrmann, Oleari,
  Remiddi, and Tausk}}]{Anastasiou:2000mf}
\bibinfo{author}{\bibfnamefont{C.}~\bibnamefont{Anastasiou}},
  \bibinfo{author}{\bibfnamefont{T.}~\bibnamefont{Gehrmann}},
  \bibinfo{author}{\bibfnamefont{C.}~\bibnamefont{Oleari}},
  \bibinfo{author}{\bibfnamefont{E.}~\bibnamefont{Remiddi}}, \bibnamefont{and}
  \bibinfo{author}{\bibfnamefont{J.~B.} \bibnamefont{Tausk}},
  \bibinfo{journal}{Nucl. Phys.} \textbf{\bibinfo{volume}{B580}},
  \bibinfo{pages}{577} (\bibinfo{year}{2000}{\natexlab{b}}),
  \eprint[http://arXiv.org/abs]{hep-ph/0003261}.

\end{thebibliography}
\end{document}